\documentclass{aa} 
\usepackage[T1]{fontenc}
\usepackage{epstopdf}
\usepackage{natbib}

\usepackage{graphicx}
\usepackage{txfonts}

\begin{document}

   \title{Is there more than meets the eye?
   Presence and role of submicron grains in debris discs}

   \author{P.Thebault
          \inst{1},
          Q.Kral\inst{1}
          }
   \institute{LESIA-Observatoire de Paris, UPMC Univ. Paris 06, Univ. Paris-Diderot, France
             }  
\offprints{P. Thebault} \mail{philippe.thebault@obspm.fr}
\date{Received ; accepted } \titlerunning{unbound grains in debris discs}
\authorrunning{Thebault \& Kral}

%

%

 
  \abstract
   {The presence of submicron grains has been inferred in several debris discs, usually because of a blue color of the spectrum in scattered light or a pronounced silicate band around 10$\mu$m, despite the fact that these particles should be blown out by stellar radiation pressure on very short timescales. So far, no fully satisfying explanation has been found for this apparent paradox. }
   {We investigate the possibility that the observed abundances of submicron grains could be "naturally" produced in bright debris discs, where the high collisional activity produces them at a rate high enough to partially compensate for their rapid removal. We also investigate to what extent this potential presence of small grains can affect our understanding of some debris disc characteristics. }
   {We use a state of the art numerical collisional code following the collisional evolution of a debris disc down to submicron grains far below the limiting blow-out size $s_{\rm{blow}}$. We consider compact astrosilicates and explore different configurations: A and G stars, cold and warm discs, "bright" and "very bright" systems. We then produce synthetic spectra and SEDs, where we identify and quantify the signature of unbound submicron grains.}
   {We find that, in bright discs (fractional luminosity $\gtrsim10^{-3}$) around A stars, there is always a high-enough amount of submicron grains to leave detectable signatures, both in scattered-light, where the disc's color becomes blue, and in the mid-IR ($10\lesssim\lambda\lesssim20\mu$m), where it boosts the disc's luminosity by at least a factor of 2 and induces a pronounced silicate solid-state band around $10\mu$m. We also show that, with this additional contribution of submicron grains, the SED can mimic that of two debris belts separated by a factor of $\sim2$ in radial distance. For G stars, the effect of $s\leq s_{\rm{blow}}$ grains remains limited in the spectra, in spite of the fact that they dominate the system's geometrical cross section. We also find that, for all considered cases, the halo of small (bound and unbound) grains that extends far beyond the main disc contributes to $\sim50$\% of the flux up to $\lambda\sim50\mu$m wavelengths.}
   {}

   \maketitle
%

\section{Introduction} \label{intro}

An important fraction of main sequence stars are known to be surrounded by debris discs, detected by the luminosity excess produced by small dust particles. In the classical view of these systems, this dust is interpreted as being produced by the steady collisional grinding of material leftover from the planet formation process. The dust-producing collisional cascade is believed to extend from planetesimal-sized parent bodies down to grains small enough to be blown out by stellar radiation pressure \citep{wyat08}. For a canonical collisional equilibrium size-distribution in $dN\propto s^{-3.5}ds$, the system's geometrical cross section is dominated by the smallest grains, of size $s_{min}$, in the cascade. These smallest grains should thus also dominate the disc's luminosity at all wavelengths shorter than $\sim 2\pi s_{min}$ \footnote{Given that, for a given wavelength $\lambda$, both the scattering and emission efficiencies quickly drop for sizes smaller than $\lambda/2\pi$ \citep[e.g.,][]{li08}}, i.e., typically from the visible up to the near-to-mid-IR. At these wavelengths observations are thus expected to be mostly sensitive to grains of sizes $s_{\rm{min}}\sim s_{\rm{blow}}$, where $s_{\rm{blow}}$ is the limiting size below which particles are blown out by stellar radiation pressure. Note that, even though detailed numerical collisional models have shown that realistic grain size distribution can significantly depart from the idealized $dN\propto s^{-3.5}ds$ power-law, they all confirm that the geometrical cross section should still be dominated by grains close to the $s_{\rm{blow}}$ value \citep{theb03,kriv06,theb07}.

However, this theoretical prediction is sometimes challenged by observations. For example, the detailed studies by \cite{pawe14} and \cite{pawe15} have shown that, for solar-type stars at least, $s_{min}$ can be significantly larger than $s_{\rm{blow}}$, with $s_{min}/s_{\rm{blow}}$ reaching values up to $\sim10$. Several explanations have been proposed for this puzzling feature, such as the non-production of very small grains resulting from the conservation of the surface energy at collisions \citep{krij14,theb16} or the imbalance between small grain production and destruction rates in dynamically "cold" discs \citep{theb08,theb16}.
For a handful of systems, however, the issue is the opposite one: observations have revealed the signature of sub-micron grains that are smaller than the blow-out size. This presence has been inferred either from a blue color in scattered light photometry, such as for the HD32297  \citep{kalas05,fitz07a}, HD15115 \citep{debes08} or AU Mic discs \citep{auge06,fitz07b,lomax18} or by strong solid-state features in the mid-IR,  such as in the HD172555  \citep{john2012b}  or HD113766 systems \citep{olof2013}  \footnote{We leave aside the probably completely different issue of exozodiacal discs for which a large population of submicron or even nano grains has been detected extremely close to their parent stars \citep[see review by][]{kral17}}.

The presence of significant amounts of $s\leq s_{\rm{blow}}$ dust is unexpected since these grains should be blown out of the system on very short timescales, of the order of one orbital period. Given this strong constraint, most explanations proposed for this puzzling presence involve a recent stochastic event, such as the catastrophic breakup of a large planetesimal \citep{john2012b,olof2013,kral15} or the powerful collisional chain reaction, also triggered by the breakup of a planetesimal, called collisional "avalanches" \citep{grig07,theb18}. However, these scenarios have the disadvantage of relying on short-lived events that are probably unlikely to be observed. 

We here reconsider this unbound grains issue within the context of standard debris discs at collisional steady state, and investigate under which circumstances such discs can "naturally" produce a steady level of small sub-micron grains that could leave a long-lived signature. There are indeed several reasons why small $s\leq s_{\rm{blow}}$ grains might, in some cases, be potentially observable in discs at collisional steady-state. 

A first potential case is that of bright and dense discs with a very high collisional activity, where we expect the depletion of grains below the $s=s_{\rm{blow}}$ limit to be significantly reduced. This is because, while the abundances of $s\geq s_{\rm{blow}}$ grains, which are produced \emph{and} destroyed in the collisional cascade, are $\propto M_{\rm{disc}}$ ($M_{\rm{disc}}$ being the total disc mass), the abundances of $s\leq s_{\rm{blow}}$ grains, which are only produced in the cascade before leaving the system, scale with $M_{\rm{disc}}^{2}$. To a first order, the density drop at the $s=s_{\rm{blow}}$ frontier should thus be $\propto 1/M_{\rm{disc}}$ and should thus decrease for discs with higher collisional activity (i.e., total mass). 
So far, the amplitude of the density drop at $s=s_{\rm{blow}}$, and in particular its link to a given star/disc configuration, has not been investigated by numerical models studying the collisional evolution of debris discs, despite of this drop being a feature that appears in most of these studies \citep{theb03,theb07,kriv06,kriv18}. To our knowledge, the only study focusing on the $s=s_{\rm{blow}}$ frontier is the one by \citet{kriv00}, which predates the advent of sophisticated collisional models and only considers the specific case of the $\beta$ Pictoris disc.

Another potential reason for the presence of submicron grains is that the simplified assumption that the ratio $\beta(s)$ between radiation pressure and gravitational forces is $\propto 1/s$, and thus keeps increasing with decreasing sizes, breaks down below a given size $s_{\rm{peak}}$ where $\beta$ reaches its maximum value. For $s\leq s_{\rm{peak}}$ grains, $\beta$ does in fact \emph{decrease} with decreasing sizes and, for solar-type stars and for most standard materials (astrosilicates, carbonaceous grains, etc...), $\beta(s)$ exceeds the limiting $\beta=0.5$ value for blowout only in a relatively narrow size range around $s_{\rm{peak}}$  \citep[see for example][]{kriv98}. Therefore, very small grains will not be blown out of the system. This explanation has been advocated by \cite{john2012b} to explain the presence of submicron grains in the HD172555 system, but only with simple qualitative arguments.

Last but not least, even if they are underabundant, small unbound grains could leave a significant signature in thermal emission because of their temperature that is generally higher than that of bigger grains closer to the blackbody value. This effect could be especially pronounced in the Wien domain of the Planck function where the emitted flux varies exponentially with the temperature.

The present paper is organized as follows. Sec.\ref{model} presents the numerical model that we use to investigate the collisional evolution of debris discs, in particular of their size distribution down to submicron grains. Sec.\ref{results} presents our main results for the different star/disc configurations that we have explored. We discuss the implications of these results in light of the observed characteristics of bright debris discs in Sec.\ref{discu}, and give our main conclusions and perspectives in Sec.\ref{ccl}.

\section{Model}\label{model}

We use an improved version of the collisional model developed by \cite{theb03} and \cite{theb07}, which is based on a statistical particle-in-a-box approach, where particles are sorted into logarithmic size bins. This code also has a 1D spatial resolution, being divided into radially concentric annuli. 

\subsection{Updated impact rate estimate procedure}

In versions of the code used so far, collision rates and impact velocities were computed using analytical equations based on the average orbital eccentricity and inclinations in each size and spatial bin (or analytical estimates of the outbound velocities for $s_{\rm{blow}}$ grains), taking into account the effect of stellar radiation pressure on small grains. 

In order to get more accurate estimates, taking for instance in account the fact that grains are not launched from perfectly circular orbits, we have changed this procedure. We now first run an $N$-body code, where we deterministically follow the dynamical evolution, taking into account the effect of stellar gravity and radiation pressure, of $\sim 10^{7}$ particles released from an extended disc of virtual parent bodies and covering the complete size range considered in the statistical code. The $N$-body code is an adaptation of the collision-tracking deterministic code developed by \cite{theb06} in the context of mutual planetesimal collisions in binary systems. All mutual close encounters are tracked and relative velocities at encounters are stored in a 5-dimension table $dv(r,s_1,s_2,r_1,r_2)$, where $r$ is the radial location of the encounter, $s_1$ and $s_2$ are the sizes of the two impacting bodies, and $r_1$ and $r_2$ the location from which they were released (which can strongly depart from $r$ for grains significantly affected by radiation pressure). This huge $dv(r,s_1,s_2,r_1,r_2)$ table is then reused in the main statistical code \footnote{ These estimated velocities do in practice never strongly depart from analytical estimates, but this numerical procedure ensures a better accuracy and dispenses with analytical calculations that can get cumbersome for non-circular launch orbits and averaging over all possible encounter angles}.

\subsection{Updated collision outcome procedure}

Collision outcomes are divided into two categories, cratering and fragmentation, depending on the ratio between the specific impact kinetic energy and the specific shattering energy $Q^{*}$, which depends on object sizes and composition. As in most similar codes, we consider two distinct prescription for $Q^{*}$:  one for the gravity regime dominating for large ($\geq 0.1-1\,$km) bodies and one for the strength regime dominating for smaller objects.

In most collisional codes, the strength regime $Q_{s}^{*}(s)$ is described by a power law with a negative exponent, using prescriptions derived from laboratory experiments \citep{housen99} or SPH simulations \citep{benz99}. 
Note, however, that these prescriptions are designed to fit a size regime that typically spans the 1\,cm-100m size domain, and that there is no guaranty that they might be extrapolated over several orders of magnitude into the micron or sub-micron size domain. As \cite{heng10} rightfully point out, the prescription \emph{has} to break down at some point because it would otherwise diverge to infinity for infinitely small bodies. \cite{heng10} and \cite{lies14} argue that the experimental results of \cite{flynn04} show that $Q_{s}^{*}(s)$ levels off to an almost constant $\sim 10^{7}$erg.g$^{-1}$ value at small target sizes. However, the \cite{flynn04} experiments only considered a very limited target size range (less than a factor 2) in the cm-domain, so that it seems difficult to conclude to a constant $Q_{s}^{*}(s)$ in the whole $s\leq 1\,$cm domain.

Unfortunately, there is, to our knowledge, no available experimental data for high-velocity collisions of micron-sized grains. In the absence of such data, one of the most useful laboratory experiments are those by \cite{naga14}, which considered impacts on targets spanning almost two orders of magnitude in sizes down to the mm domain, and at impact velocities up to $\sim 1\,$km.s$^{-1}$. We consider the results displayed in Fig.7 of that paper as a reference for deriving a power law prescription for $Q_{s}^{*}(s)$, which we then extrapolate down to the sub-micron domain and up to the transition size where gravity takes over. For the gravity regime, we still follow the prescription by \cite{benz99}. We also include a simplified dependence on impact velocity derived from \cite{stew09} so that our full $Q$*(s) prescription reads
\begin{equation}
Q^{*} = Q_{0} \left(\frac{s}{1\rm{cm}}\right)^{-0.4} \left(\frac{v_{\rm{coll}}}{v_{1}}\right)^{0.5} + B\,\rho \left(\frac{s}{1\rm{cm}}\right)^{1.36} \left(\frac{v_{\rm{coll}}}{v_{2}}\right)^{0.5} 
\label{Qs}
\end{equation}
wtih $Q_0=7\times10^{6}$erg.g$^{-1}$, $v_1$=2.5$\times10^{4}$cm.s$^{-1}$ is the typical velocity of the \cite{naga14} experiments, $v_2=3\times10^{5}$cm.s$^{-1}$ and $B=0.3\,$erg.cm$^{3}$.g$^{-2}$.

For both fragmenting and cratering regimes, the size of the largest fragment and the size distribution of the other debris are then derived through the energy scaling prescriptions presented in \citet{theb03} and  \citet{theb07}. Following \cite{theb16}, we also implement a surface energy-conservation criterion derived from \citet{krij14} that gives the lower limit for fragment sizes after any collision.

\subsection{Setup}\label{setup}

\begin{table}[h]
\caption{Set-up summary}
\centering
\begin{tabular}{lc}
\hline\hline

Stellar-type & A6V or G2V \\
\hline
Material & Compact astrosilicate \\
Blow-out size ($s_\mathrm{\rm{blow}}$) & $1.6\,\mu$m (A6V star)\\
 & $0.4\,\mu$m (G2V star)\\
Minimum particle size & $ 0.02\,\mu$m  \\
Maximum particle size   & 2\,km  \\
Initial size distribution & $\textrm{d}N \propto s^{-3.5} \textrm{d}s$ \\
Dynamical excitation & <e>=2<i>=0.075\\
Radial extent / cold disc case & 50-90\,au (A6V star) \\
  & 30-50\,au (G2V star) \\
Radial extent / warm disc case & 5-9\,au (A6V star) \\
  & 3-5\,au (G2V star) \\
fractional luminosity at steady state & $f_{d}\sim10^{-3}$ ("bright" disc)\\
 & $f_{d}$$\sim$5$\times10^{-3}$ ("very bright" disc)\\
\hline
\end{tabular}
\label{setupt}
\end{table}

We consider the most generic possible case of a "classical" debris disc at collisional equilibrium and explore 3 crucial parameters for the problem at hand: stellar type, disc location and total disc mass. In order to keep the parameter exploration to a manageable level, we consider 2 representative configurations for each of these parameters. For the stellar type, we consider a $\beta$-Pic like A6V type and a solar-type G2V star. For disc location, we consider the case of a cold and a warm belt. The cold belt extends from $r_{\rm{in}}=50$ to $r_{\rm{out}}=90\,$AU for the A6V case and from 30 to 50\,AU (roughly matching a Kuiper Belt) for the G2V star. As for the warm belt, it extends from 5 to 9\,au for the A6V case and from 3 to 5\,AU (roughly matching an asteroid belt) for the G2V star. Note that these radial extents are those of the main "parent body disc" where all the mass is initially located, but that we also simulate the evolution of the much more extended halo of small grains produced in this main disc but pushed to larger radial distances by radiation pressure (either on highly eccentric bound orbits or by direct blow-out on unbound trajectories).

As for the total disc mass, we chose to parameterize it by the disc's IR fractional luminosity $f_d$. Since the presence of unbound grains is expected to be favoured in bright and massive discs (see Sec.\ref{intro}), we consider the case of a typical "bright" disc with $f_d\sim 10^{-3}$, similar to the archetypal $\beta$--Pictoris system \citep{decin00} and a "very bright" disc with $f_d\sim 5\times10^{-3}$, such as the HR4796  \citep{jura98} or HD32297 \citep{moor06} systems. Note that these $f_d$ values are not necessarily the ones at t=0 but rather the ones once a steady-state is reached. In practice, we always start with discs with $f_d$ larger than the one we are aiming for, and let the systems collisionally evolve until 1) the shape of the particle size distribution (PSD) no longer changes (steady-state), and 2) $f_d$ decreases to the desired value.
For the disc's dynamical excitation, we chose a "standard" configuration with <e>=2<i>=0.075 \citep{theb09}.

\begin{figure}
\includegraphics[scale=0.5]{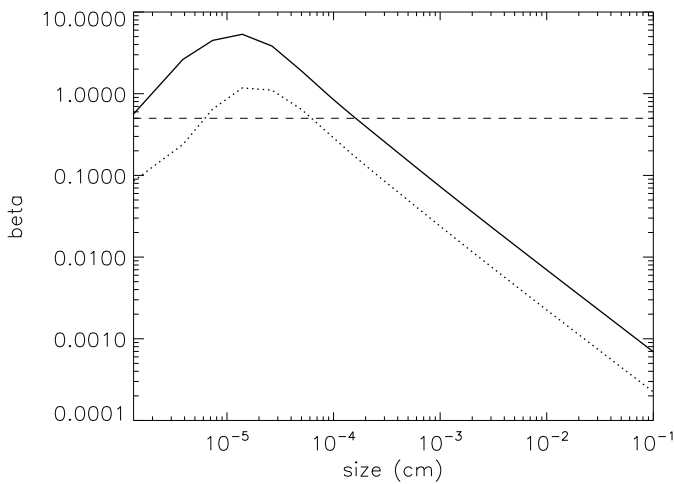}
\caption[]{Value of $\beta$, the ratio between radiation pressure and gravitational forces, as a function of particle size, computed for non-porous astrosilicates. The $\beta$(s) curve is shown for a luminous A6V star (solid) as well as for a solar type G2V star (dotted). The dashed horizontal line marks the $\beta=0.5$ limit, above which grains are no longer on bound orbits around the star (if produced from progenitors on circular orbits).}
\label{beta}
\end{figure}

Given the specificity of the problem we are addressing, we consider the evolution of the PSD from $s_{max}=2\,$km down to a very small size $s_{min}=0.02\mu$m. As for dust material, we consider a standard case of compact astrosilicates \citep{drai03}. Fig.\ref{beta} displays the $\beta$(s) curves for this material and the two stellar types considered here.

All main parameters are summarized in Tab.\ref{setupt}

\section{Results}\label{results}

\begin{figure}
\includegraphics[scale=0.5]{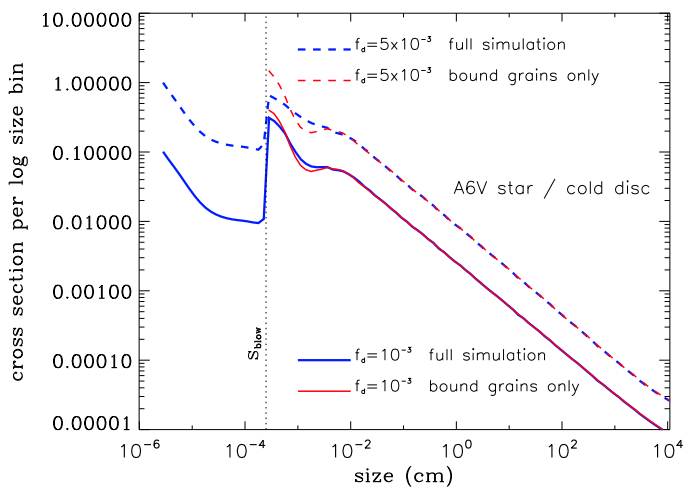}
\caption[]{A6V star, "cold" disc ($50\leq r \leq 90\,$au): Normalized disc-integrated particle size distribution, at collisional steady state, for a "bright" debris disc with $f_d\sim10^{-3}$,  and a "very bright" debris disc with $f_d\sim 5\times10^{-3}$. The red line corresponds to a simulation only taking into account $s\geq s_{\rm{blow}}$ grains. The vertical dotted line corresponds to the blow-out size $s_{\rm{blow}}$.}
\label{psdbpcoldsil}
\end{figure}

\begin{figure*}
\makebox[\textwidth]{
\includegraphics[scale=0.5]{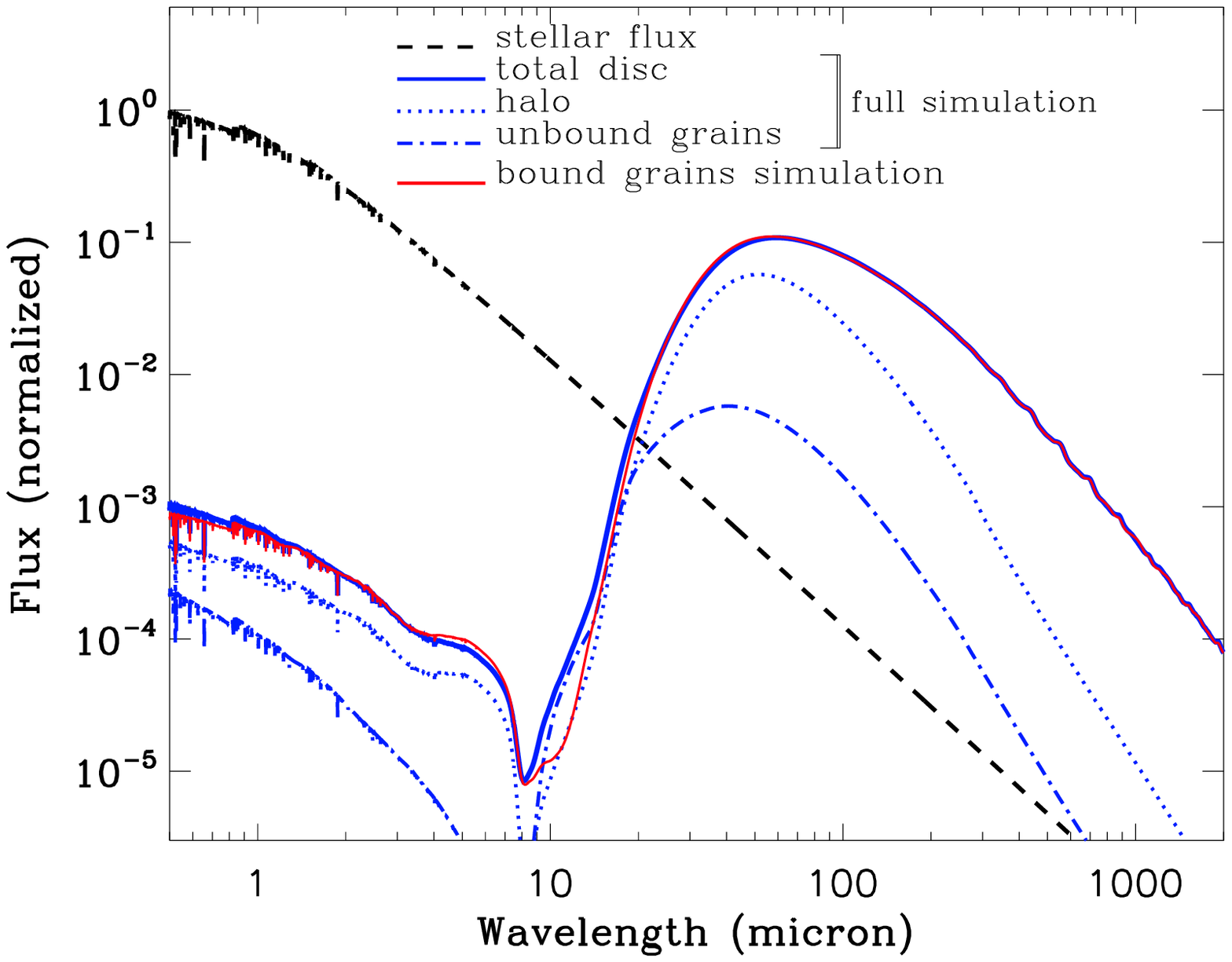}
\includegraphics[scale=0.5]{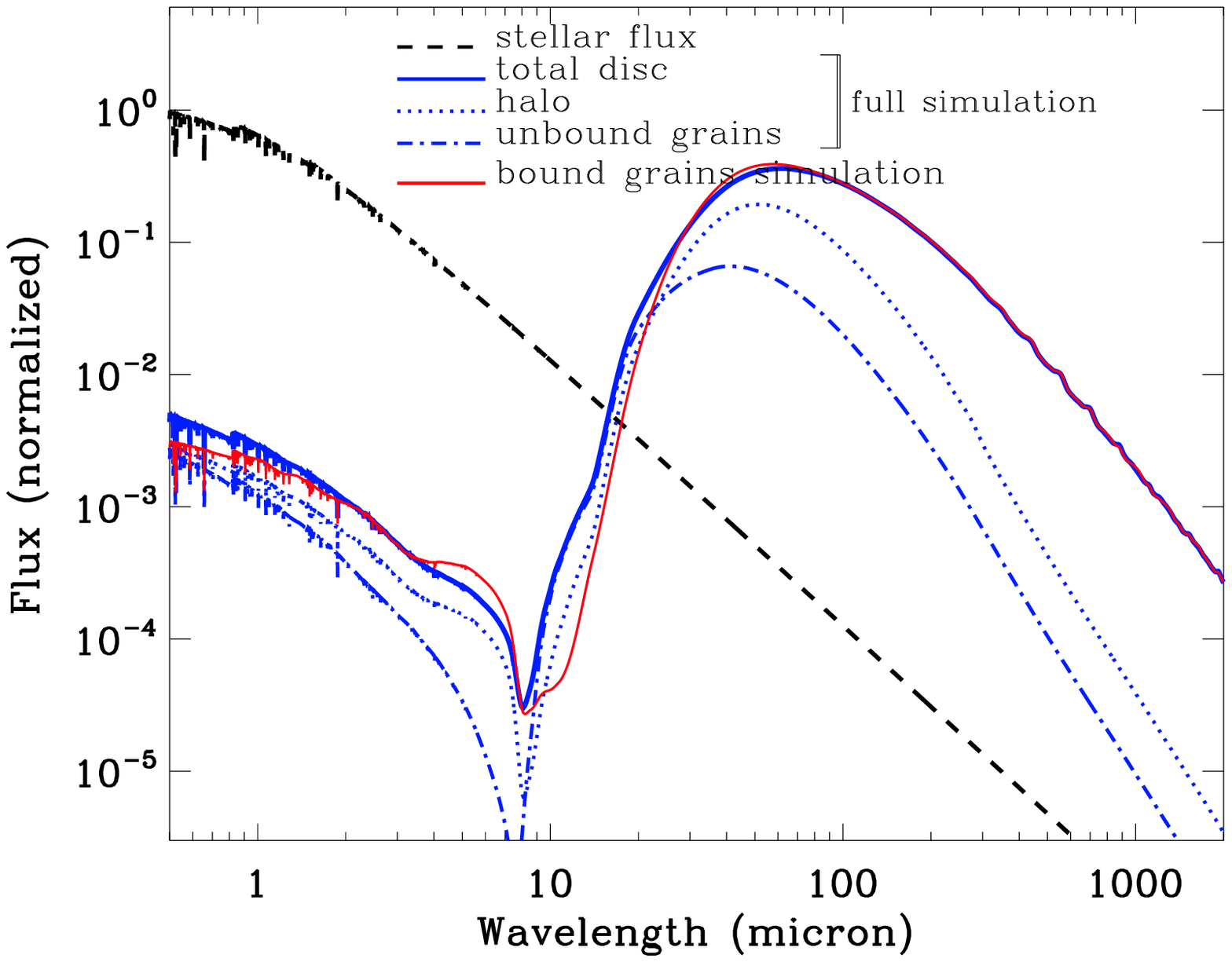}
}
\caption[]{A6V star, "cold" disc : Normalized SED (solid line in blue). \emph{left}: "bright" debris disc with $f_d\sim 10^{-3}$.  \emph{right}: "very bright" debris disc with $f_d\sim 5\times10^{-3}$. The dash-dotted blue line gives the contribution of unbound  $s\leq s_{\rm{blow}}$ grains. The dotted blue line gives the contribution of grains in the "halo" beyond the main disc. The red line corresponds to a simulation only taking into account $s\geq s_{\rm{blow}}$ grains. }
\label{sedbpcoldsil}
\end{figure*}

\begin{figure*}
\makebox[\textwidth]{
\includegraphics[scale=0.5]{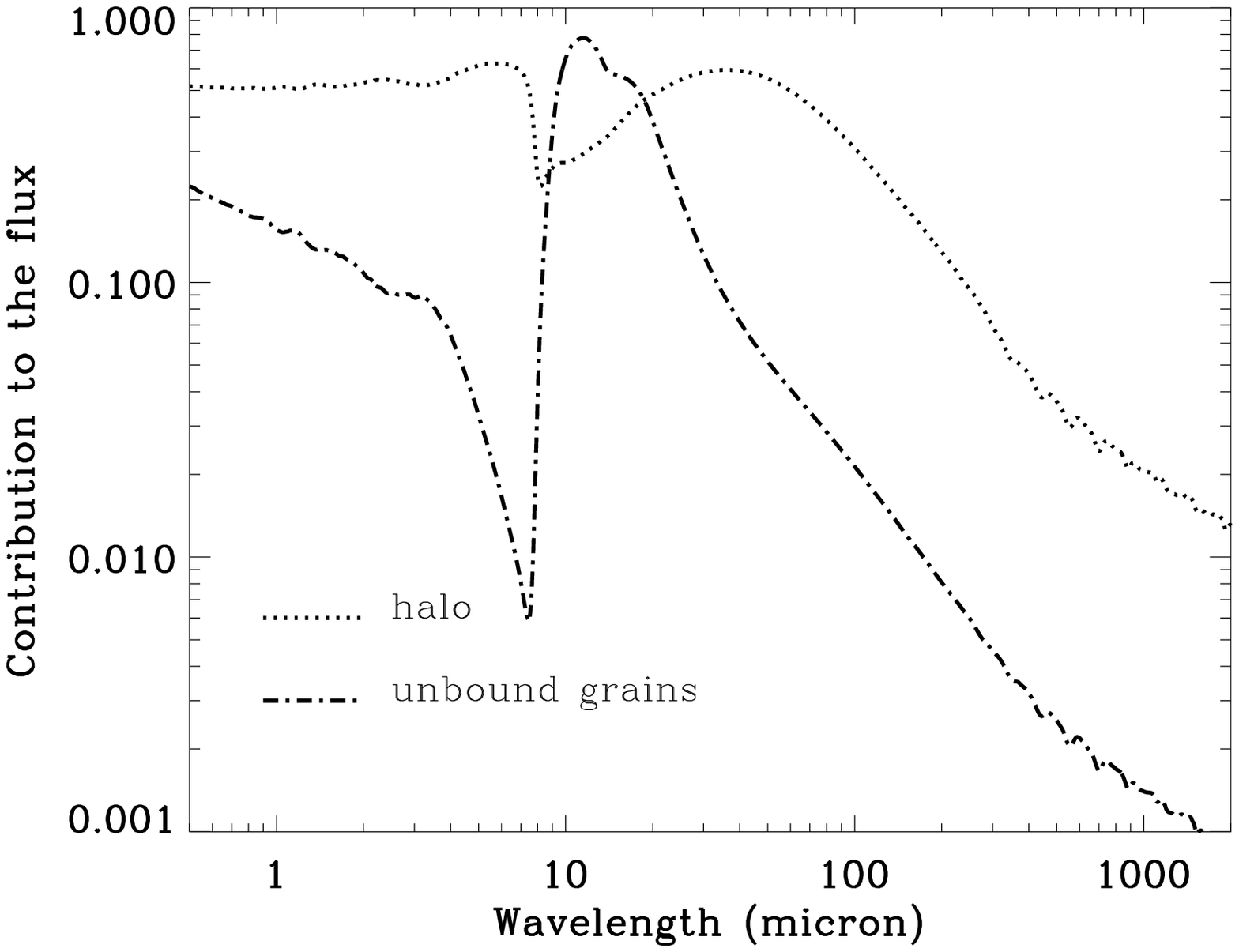}
\includegraphics[scale=0.5]{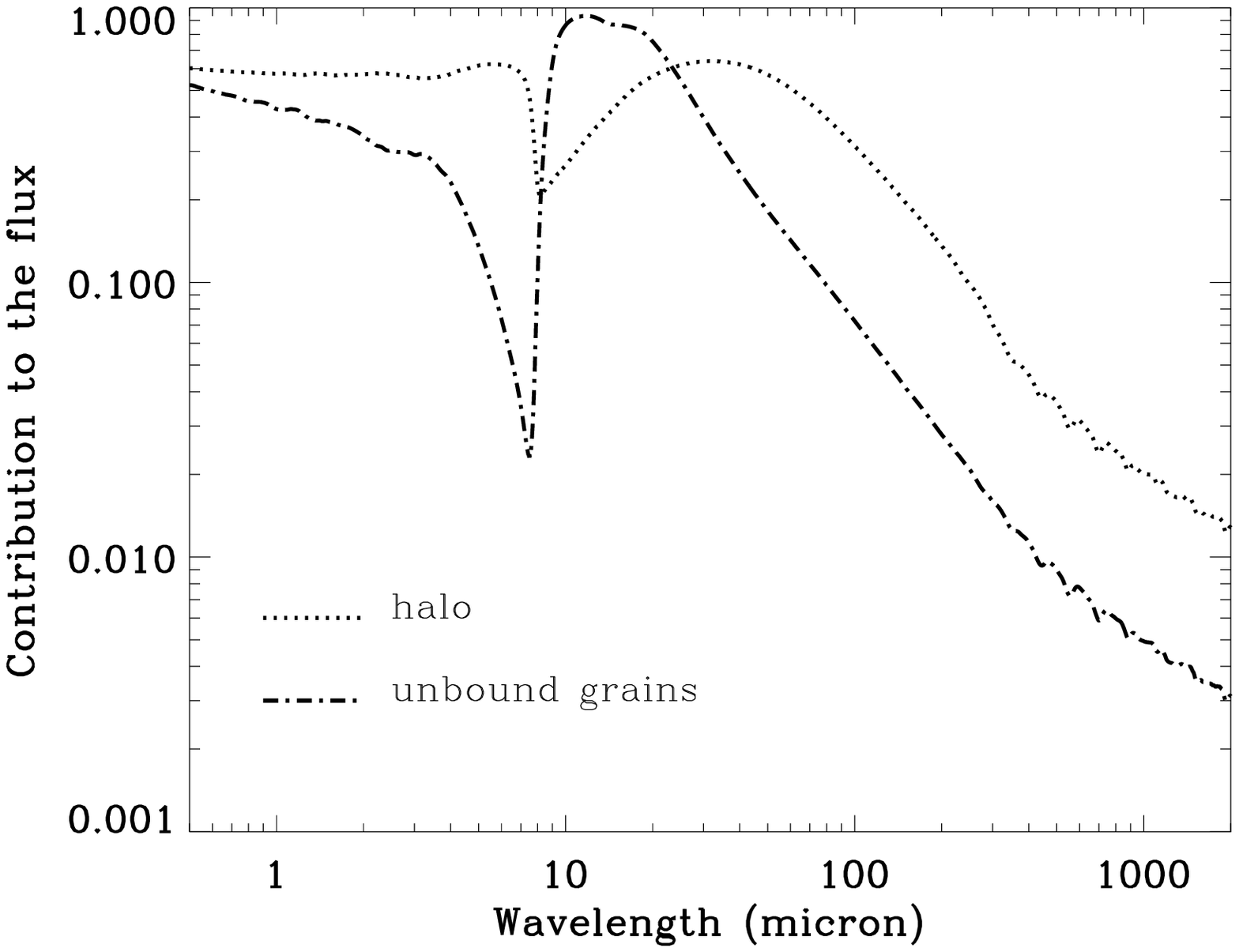}
}
\caption[]{A6V star, "cold" disc: Fractional contribution of unbound (dash-dotted line) and halo (dotted line) grains to the SED as a function of wavelength. \emph{left}: "bright" debris disc with $f_d\sim 10^{-3}$.  \emph{right}: "very bright" debris disc with $f_d\sim 5\times10^{-3}$.}
\label{fracbpcoldsil}
\end{figure*}

\subsection{A type star}

\subsubsection{Cold outer belt}

Fig.\ref{psdbpcoldsil} displays the particle size distribution (PSD) at collisional steady-state for both a "bright" and a "very bright" disc. The PSDs show well known features, such as a power-law distribution in $dN\propto s^{-q}ds$ that holds in most of the bound-grains size domain, with $q$ slightly exceeding the canonical 3.5 value because of the size-dependence of the $Q_{s}^{*}(s)$ prescription \citep[][and references therein]{gasp12,kral13}. We also see the well known "waviness" in the PSD at the lower end of the bound-grains size domain, which is due to the discontinuity at the $s=s_{\rm{blow}}$ frontier \citep{camp94,theb03,theb07}.

If we now focus on the unbound grains domain, we observe, as expected, a drop at the $s=s_{\rm{blow}}$ transition. For the $f_d=10^{-3}$ disc, this density drop is very pronounced, close to two orders of magnitude, so that, even if the PSD increases again for grains in the $\simeq  0.1\mu$m range, the disc's total geometrical cross section is largely dominated by bound grains. In contrast, for the "very bright" $f_d=5\times10^{-3}$ disc case, the density drop around $s_{\rm{blow}}$ is much less pronounced, of the order of a factor 5 only. As a consequence, the upturn in the PSD in the $s\leq0.1\mu$m domain is able to make the smallest grains the dominant contributors to the total cross section (Fig.\ref{psdbpcoldsil}). However, this might not necessarily translate into a significant contribution to the observed flux, because these grains might have a low emissivity due to their very small sizes. 

To check this, we use the GraTeR package \citep{auge99} to derive synthetic SEDs (Fig.\ref{sedbpcoldsil}). We also estimate, at all wavelengths, the relative contribution of unbound grains to the disc's total flux (Fig.\ref{fracbpcoldsil}). In order to simulate the case of an unresolved disc, we take into account the contribution of grains in the extended "halo" located beyond the main disc, which have been collisionally produced in the main belt but populate the $r\geq r_{\rm{out}}$ region because radiation pressure places them on unbound or high-eccentricity orbits \citep{stru06,theb08}. We see that, even for the $f_d=10^{-3}$ disc case where they do not dominate the cross section, unbound grains can have a significant contribution to the SED, exceeding 50\% of the flux, albeit in a relatively narrow domain between $\lambda\sim10$ and $\sim20\,\mu$m (Fig.\ref{fracbpcoldsil}a). This is partly due to the well known silicate band around $\sim10\,\mu$m (see Sec.\ref{silband}). However, the dominant effect is that the $10\leq\lambda\leq20\,\mu$m domain is a "sweet spot" where the higher temperature of these submicron grains, which exceeds the almost blackbody temperature of larger particles, has a dramatic effect. This is because, at $\sim70\,$AU from an A6V star, the $\lambda\leq20\,\mu$m domain is deep in the Wien side of the Planck function, for which the flux increases exponentially with temperature. This exponential dependence is strong enough to compensate for the lower emissivity of grains that are smaller than the $s=\lambda/2\pi$ limit. Beyond $\lambda\sim20\mu$m, we leave this purely exponential temperature-dependence regime, and the higher temperature of the tiniest grains can no longer mitigate their decreasing emissivity. Below $\lambda\sim10\mu$m the flux becomes dominated by scattered light, for which the contribution of unbound grains sharply drops before increasing again to $\sim$20\% of the flux at $\lambda\sim0.5\mu$m (Fig.\ref{fracbpcoldsil}a).

For a very bright disc with $f_d\sim 5\times10^{-3}$, the contribution of unbound grains is, as expected, even more important. They contribute to more than 50\% of the flux in the $8\leq\lambda\leq30\,\mu$m domain, and even up to 80--90\% for $10\leq\lambda\leq20\,\mu$m (Fig.\ref{fracbpcoldsil}b). Moreover, they also contribute to more than 30\% of the scattered-light flux at all wavelengths shorter than $\sim3\mu$m.

Unbound grains have also an indirect way of affecting the disc's evolution, by collisonally eroding the population of small bound particles above the $s_{\rm{blow}}$ limit, an effect first noted in the avalanche simulations of \cite{theb18}. To assess the importance of this effect we performed additional simulations where only bound $s\geq s_{\rm{blow}}$ grains were taken into account and compare the obtained PSDs and SEDs to those obtained for our full simulations \footnote{For these "bound-only" simulations, we take the same number density of large parent bodies ($s>1\,$m), i.e., the same disc mass, as for the corresponding full-simulation. As a result, the steady-state $f_d$ of these bound-only systems can in principle differ from the one of the full-simulation because of the absence of unbound grains and different number densities of particles close to the $s_{\rm{blow}}$ limit. However, despite significant flux differences in some specific narrow wavelength domains (for instance around $\lambda\sim$10-20$\mu$m), the differences in global $f_d$ always remain relatively limited (less than $\sim$25\%), mainly because the drop in $f_d$ due to the absence of unbound dust is to a large extent compensated by the $f_d$ increase due to the excess of small-bound grains that are not destroyed by these absent submicron grains. For the sake of simplicity and readability, we thus chose to label both the "full" simulation and its "bound-only" counterpart with the $f_d$ value of the full simulation.}.
For the bright $f_d=10^{-3}$ disc, the effect is already noticeable on the PSD, but remains marginal. It is for the very bright disc case that the eroding effect leaves a very pronounced signature on the PSD, with a deficit of at least a factor 2 of small bound grains in the full-simulation as compared to the bound-grains-only run (Fig.\ref{psdbpcoldsil}).

Another important result is that, regardless of the disc's mass or brightness, the halo beyond the main belt has a significant contribution to the flux, of the order of $\sim50$\%, at almost all wavelengths $\lambda\lesssim50\mu$m (Fig.\ref{fracbpcoldsil}). Its contribution then drops beyond $\lambda\sim50\mu$m because it does only contain grains smaller than $s\sim$10--20$\mu$m, which are poor emitters at these long wavelengths.

\subsubsection{Warm inner belt}

\begin{figure}
\includegraphics[scale=0.5]{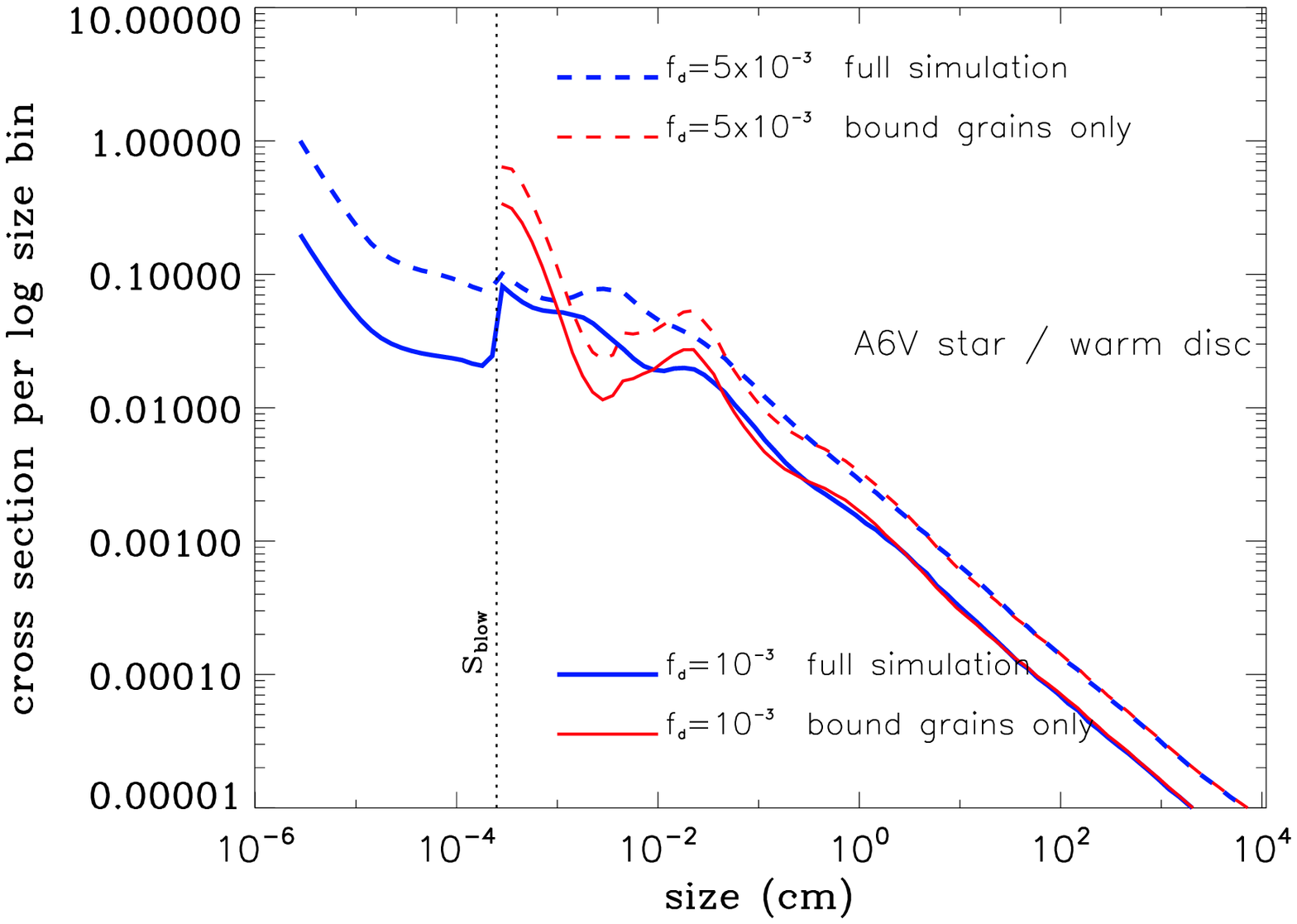}
\caption[]{Same as Fig.\ref{psdbpcoldsil}, but for a "warm" disc ($5\leq r \leq 9\,$au).}
\label{psdbphotsil}
\end{figure}

\begin{figure*}
\makebox[\textwidth]{
\includegraphics[scale=0.5]{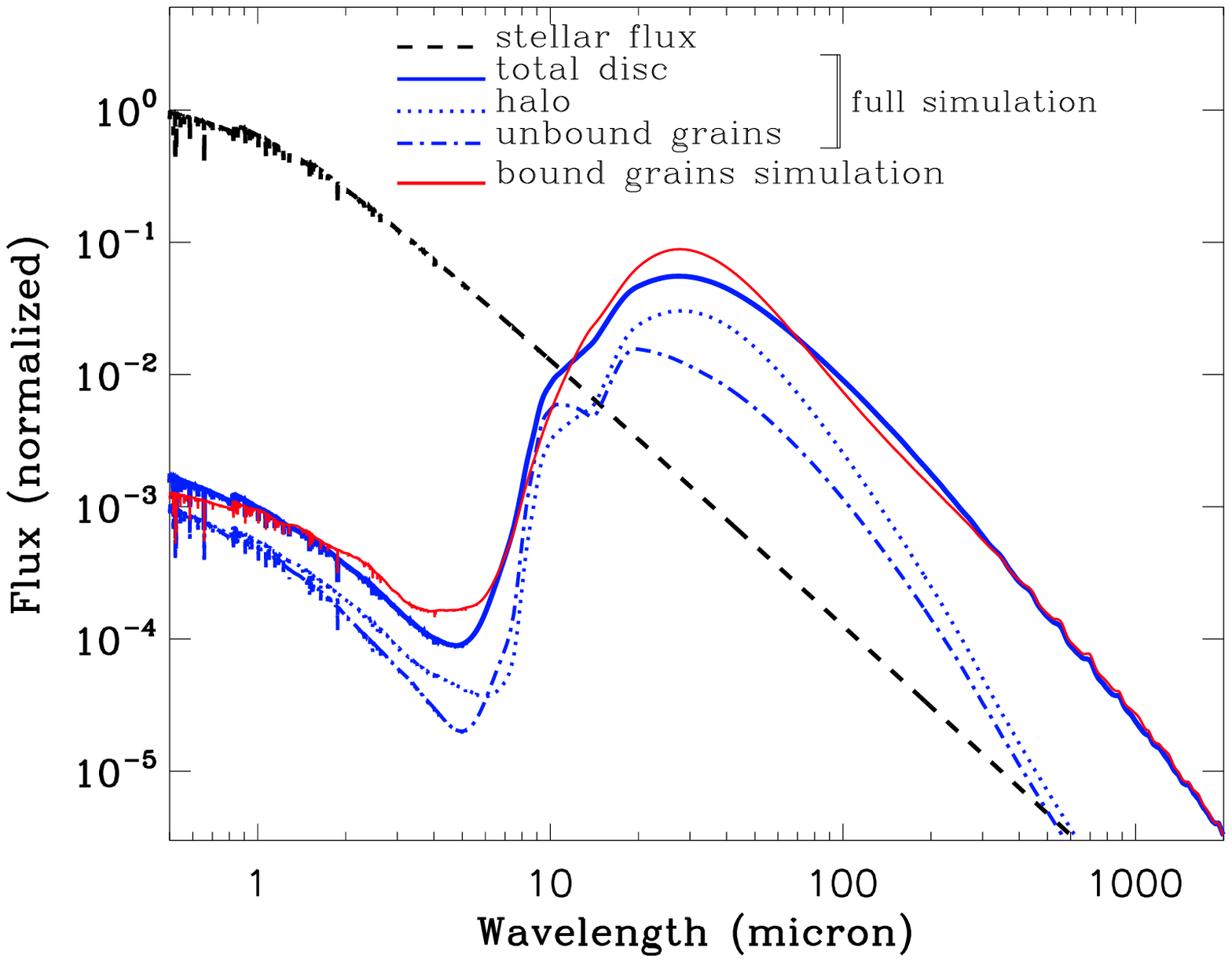}
\includegraphics[scale=0.5]{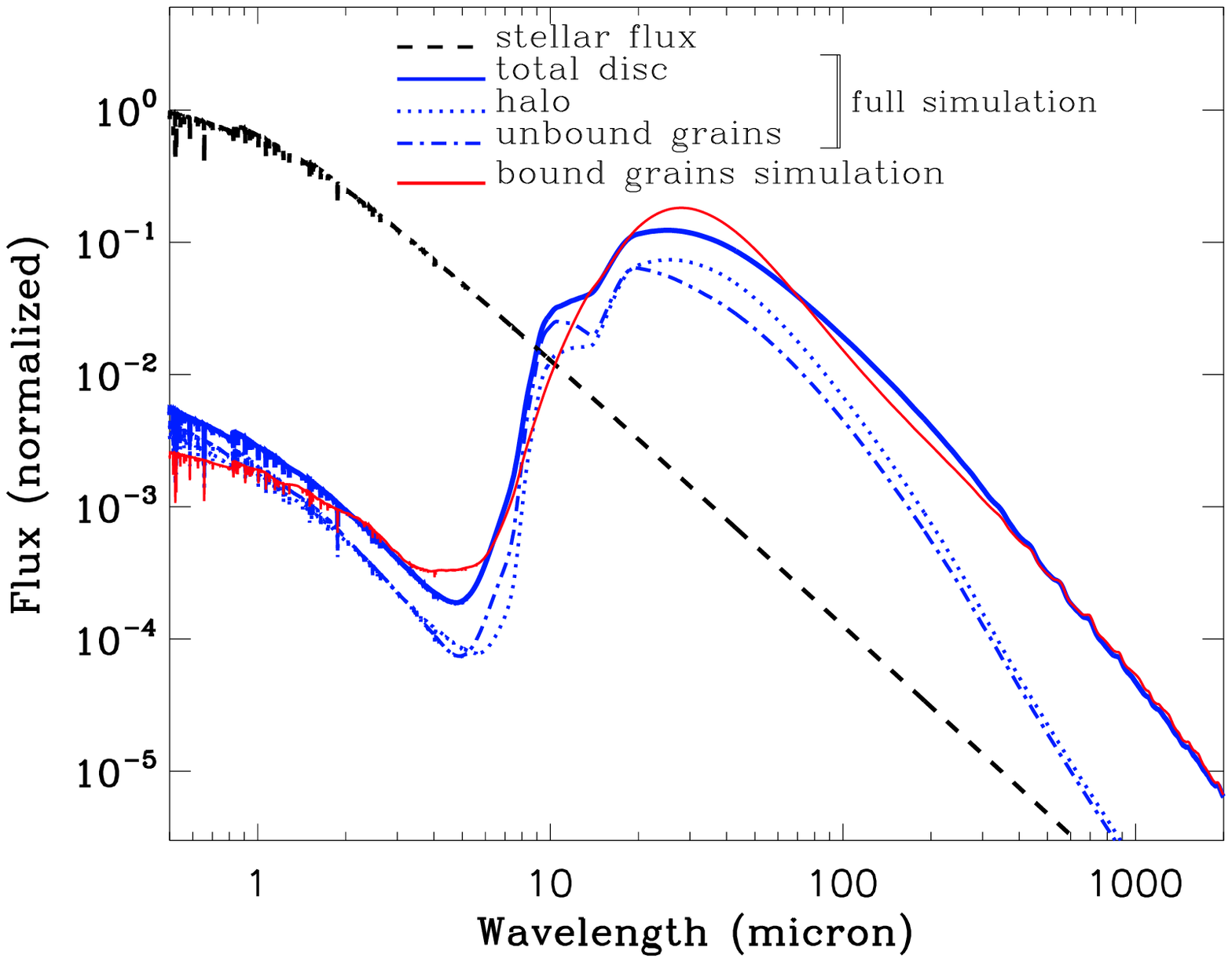}
}
\caption[]{Same as Fig.\ref{sedbpcoldsil}, but for a "warm" disc ($5\leq r \leq 9\,$au) with $f_d=10^{-3}$ (left panel) and $f_d=5\times10^{-3}$ (right panel).}\label{sedbphotsil}
\end{figure*}

\begin{figure*}
\makebox[\textwidth]{
\includegraphics[scale=0.5]{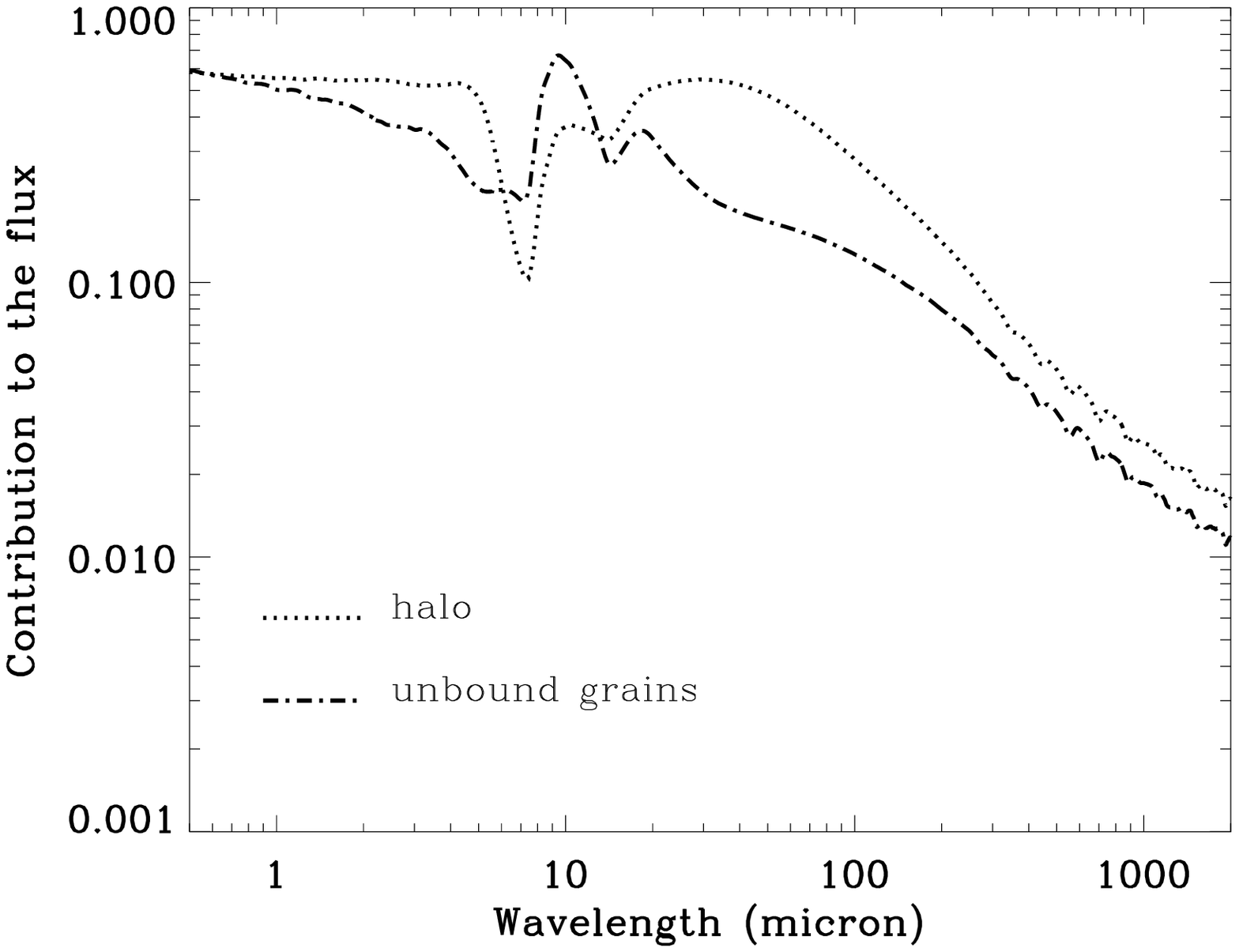}
\includegraphics[scale=0.5]{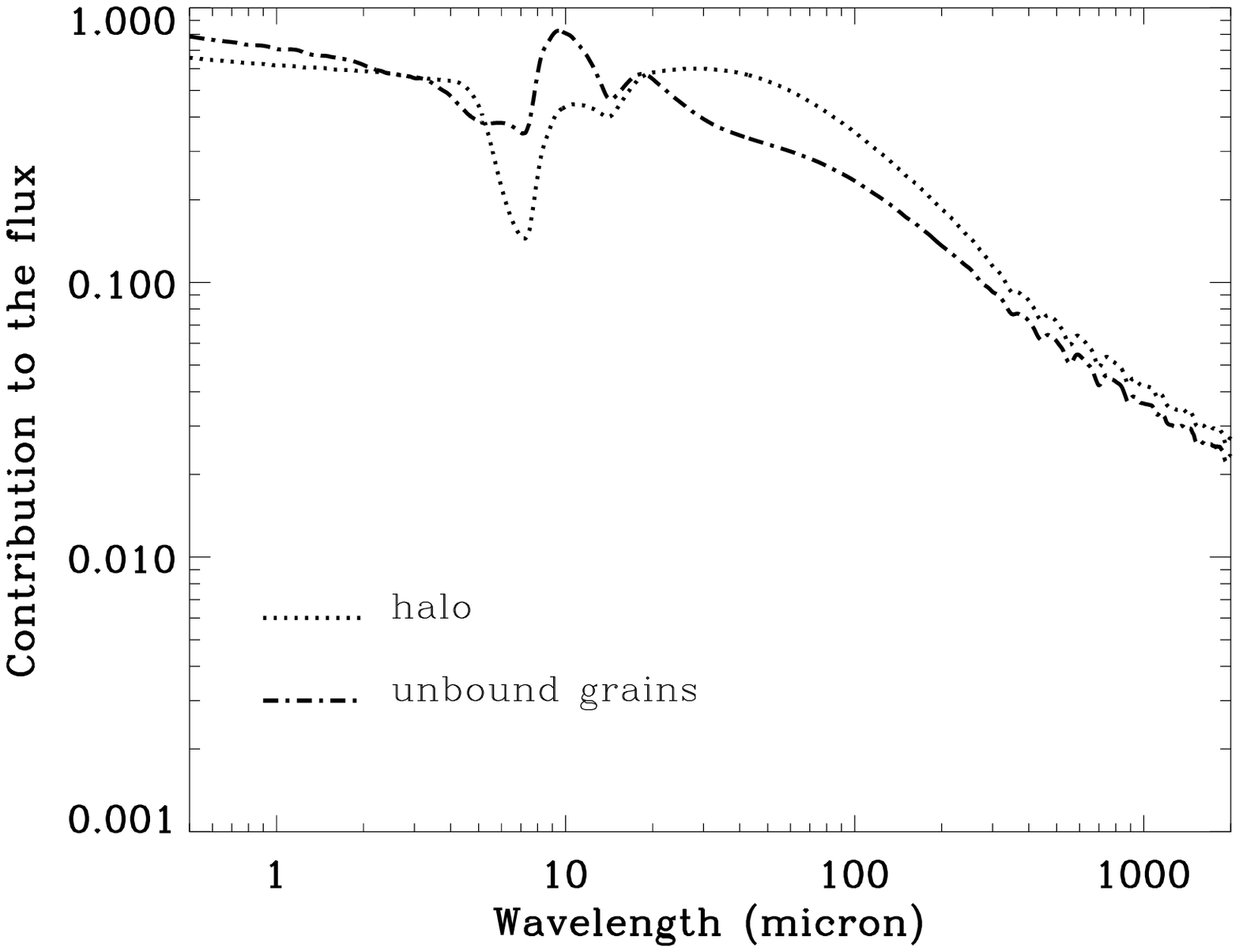}
}
\caption[]{Same as Fig.\ref{fracbpcoldsil}, but for a "warm" disc ($5\leq r \leq 9\,$au) with $f_d=10^{-3}$ (left panel) and $f_d=5\times10^{-3}$ (right panel).}
\label{fracbphotsil}
\end{figure*}

The effect of unbound grains on the PSD is stronger when considering an asteroid-like warm disc ($5\leq r\leq9\,$au). Unbound grains leave a strong signature even for a disc of fractional luminosity $f_d\sim10^{-3}$, for which we see (Fig.\ref{psdbphotsil}) that the collisional cross section is clearly dominated by $s\leq s_{\rm{blow}}$ grains. We also see that the collisional erosion of small bound particles due to these unbound grains is more efficient than in the cold disc case, and is able to reduce the number density of $s_{\rm{blow}}\leq s \leq 2s_{\rm{blow}}$ grains by a factor of $\sim5$. As expected, the effect of unbound grains becomes even more important for a brighter $f_d=5\times10^{-3}$ disc, for which the density drop at the $s_{\rm{blow}}$ frontier is almost completely erased.
The explanation for this enhanced effect of unbound grains for the same $f_d$ is that impact velocities, and thus the collisional activity, are here higher than in a cold disc located beyond 50\,au \footnote{The absolute collision rates are also higher, for the same $f_d$, than for a cold disc. However, what matters for the amplitude of the drop at $s_{\rm{blow}}$ is the \emph{ratio} between the collisional and the dynamical (i.e., orbital) timescales, and this ratio is inversely proportional to the normal geometrical depth $\tau$, and thus to $f_d$ \citep{zuck12}}.

Not surprisingly, the unbound grains also leave a significant imprint on the SEDs (Fig.\ref{sedbphotsil}), albeit in a different way as for the cold disc case. The peak due to the silicate band at $\sim10\mu$m is here much more prominent, and clearly dominates over the thermal continuum contribution in the wavelength domain. This is both because the peak's amplitude is directly proportional to the unbound grains' number density, which is higher than in the cold disc case, and because the Wien continuum in the $\lambda\sim$10-15$\mu$m domain is less steep for this warmer case, making the feature easier to discern. The depletion of small bound grains in the $s\sim1$--5$\mu$m size domain results in a flux deficit in the $\lambda\sim15$--40\,$\mu$m range, which, combined with the aforementioned bump around $\lambda\sim8-15\,\mu$m, leaves a much flatter SED in the whole $\lambda\sim8$--40$\,\mu$m domain as compared to the bound-grains-only simulation (red curve on Fig.\ref{sedbphotsil}).

\subsection{Solar-type star}

Considering a solar-type star changes two important things. Firstly, the blowout size is smaller, around $s_{\rm{blow}}\sim0.4\mu$m, and, secondly, there is a second critical "blowout" size $s_{blow2}\sim0.07\mu$m \emph{below} which particles' orbits become bound again (Fig.\ref{beta}).

\subsubsection{Cold outer belt}

\begin{figure}
\includegraphics[scale=0.5]{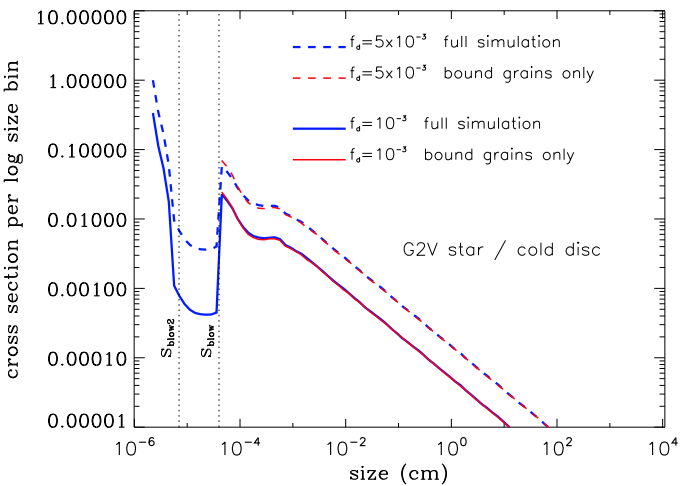}
\caption[]{same as Fig.\ref{psdbpcoldsil}, but for a solar-type G2V star. The vertical dotted lines correspond to the 2 blow-out size limits $s_{blow}$ and $s_{blow2}$}
\label{psdsun}
\end{figure}

\begin{figure*}
\makebox[\textwidth]{
\includegraphics[scale=0.5]{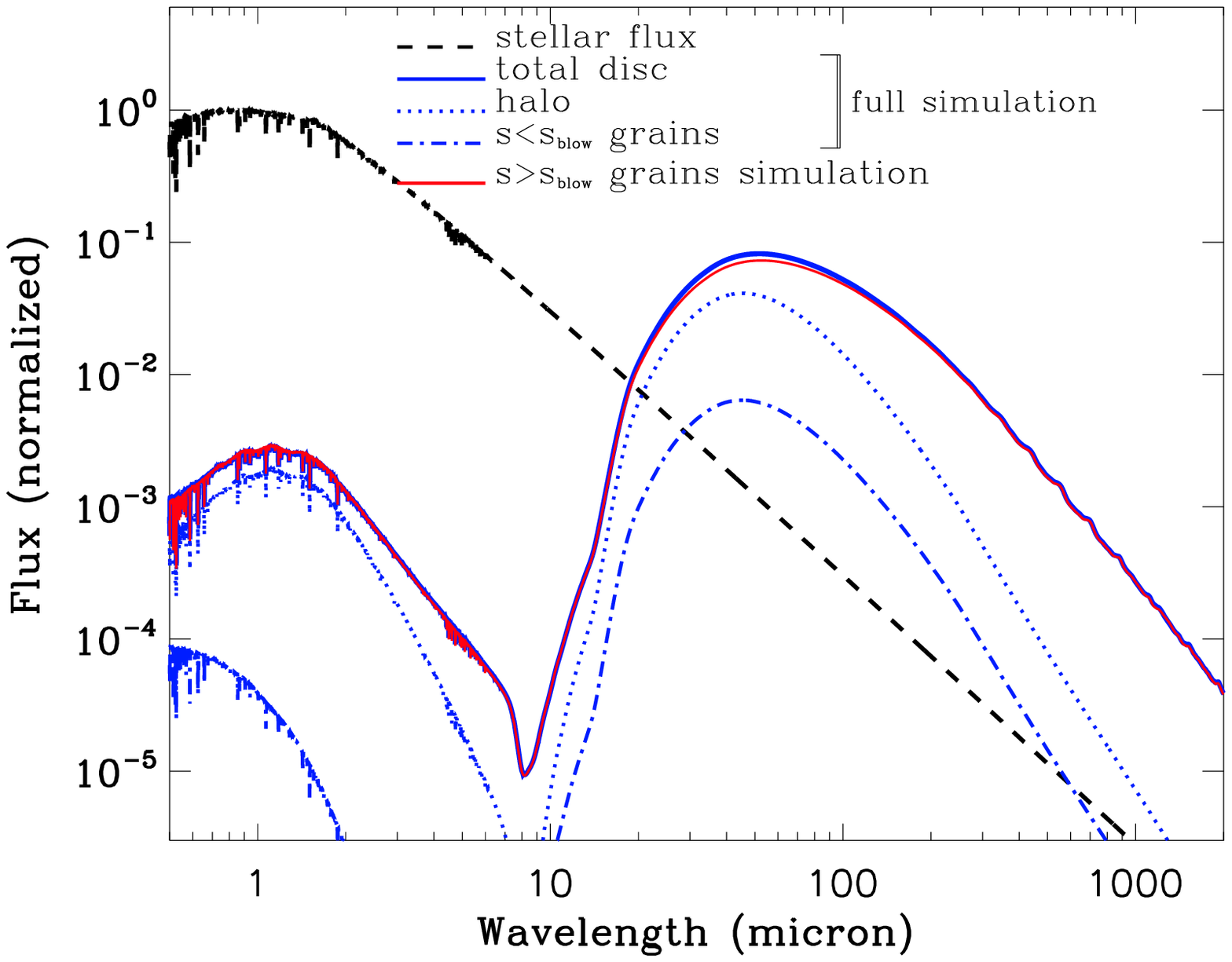}
\includegraphics[scale=0.5]{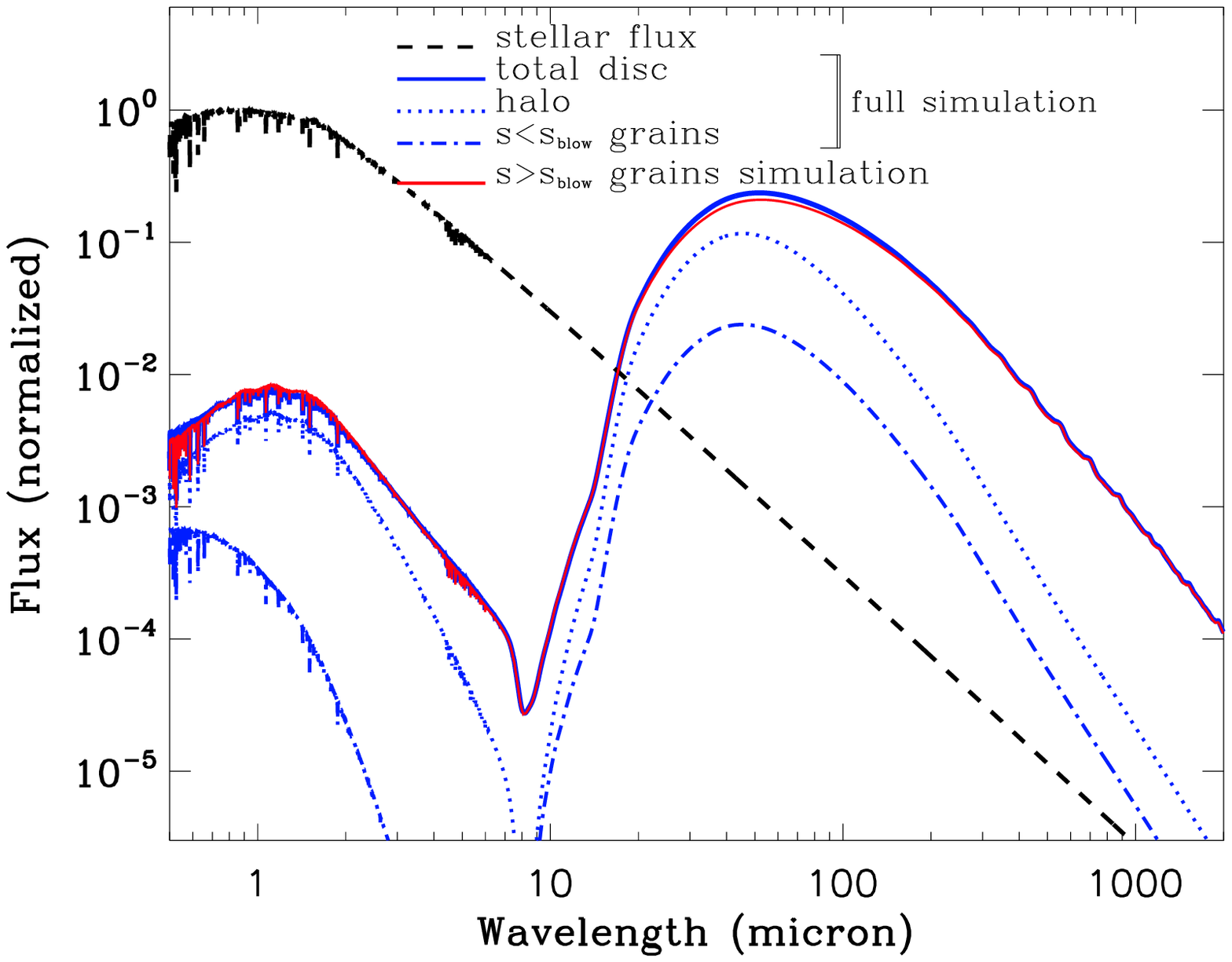}
}
\caption[]{same as Fig.\ref{sedbpcoldsil}, but for a solar-type G2V star with $f_d=10^{-3}$ (left panel) and $f_d=5\times10^{-3}$ (right panel)}
\label{sedsun}
\end{figure*}

As can be seen on Fig.\ref{psdsun}, the PSD is schematically divided into 3 different domains: a classical power-law (with a slight waviness at its lower end) profile down to  $s_{\rm{blow}}$, followed by a sharp drop and an almost plateau value in the $s_{blow2}\leq s\leq s_{\rm{blow}}$ interval, and, finally, a very strong upturn for $s\leq s_{blow2}$ grains due to the fact that these grains are no longer blown out of the system and stay on bound orbits. This upturn is strong enough for the system's total cross section to be dominated, at more than 90\%, by these $s\leq s_{blow2}$ particles, and this even in the $f_d=10^{-3}$ disc case. However, this vast population of tiny grains has a very limited erosion effect on the population of small bound grains, with the $s\geq s_{\rm{blow}}$ PSD staying very close to what it is in a bound-grains-only simulation. This is because these tiny grains' bound orbits make them impact larger grains at much smaller velocities than in the case of an A star that blows them away on high-speed outbound trajectories.

Note also that the $s\leq s_{blow2}$ grains dominance over the cross section does not translate into a significant effect on the SED (Fig.\ref{sedsun}), with a contribution that almost never exceeds 10\% of the flux at any wavelengths, even for a dense very bright $f_d=5\times10^{-3}$ disc. The higher temperature of these grains is here not able to compensate for their extremely small emission coefficient due to their very small size. While the silicate band is clearly noticeable around 11$\mu$m, it is mostly due to $s\geq s_{\rm{blow}}$ grains, because the lower value of the $s_{\rm{blow}}$ limit allows for sub-micron grains to contribute to it even in a bound-grains only simulation.

\subsubsection{Warm inner belt}

\begin{figure}
\includegraphics[scale=0.5]{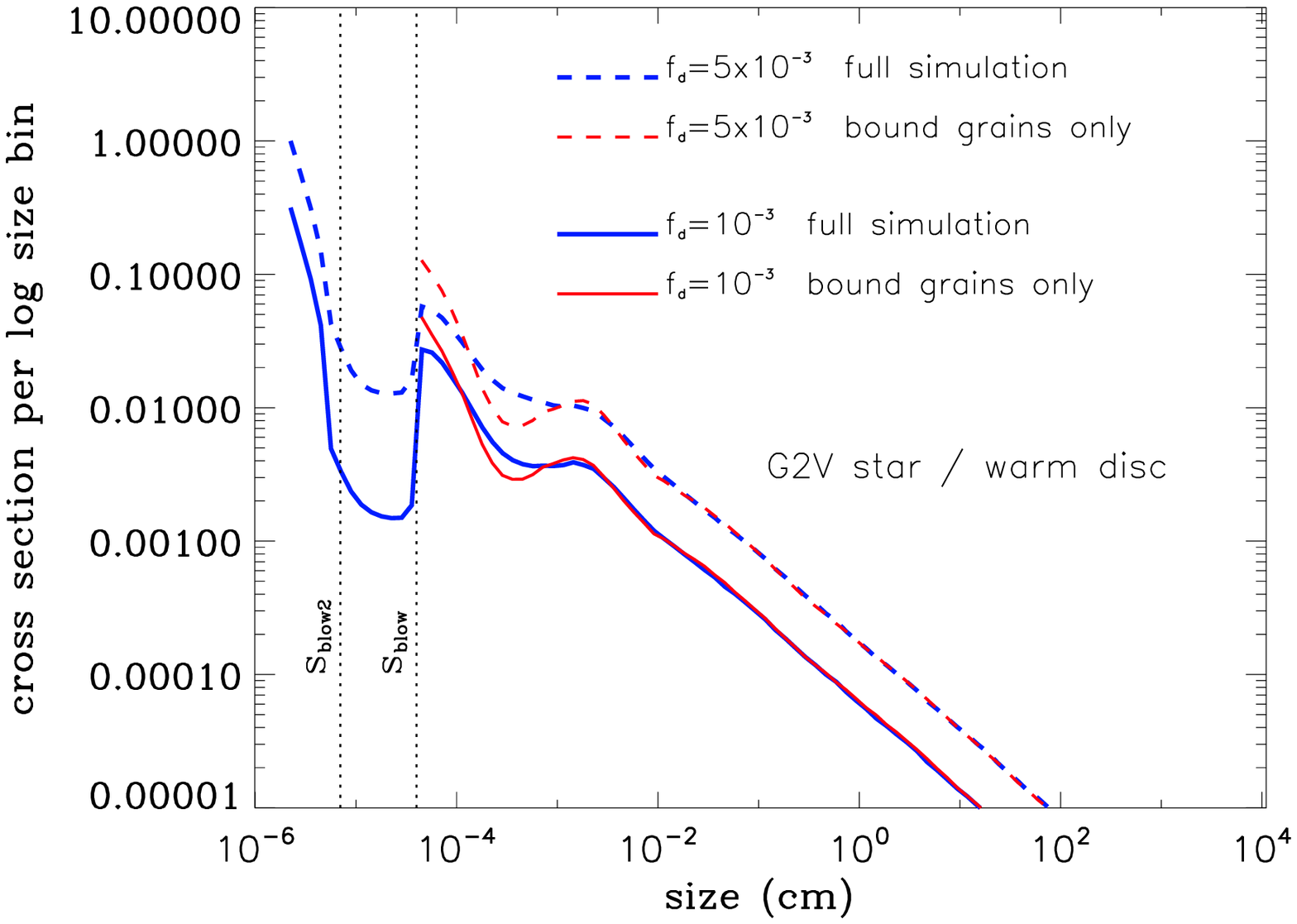}
\caption[]{Same as Fig.\ref{psdbpcoldsil}, but for a solar-type G2V star. The vertical dotted lines correspond to the 2 blow-out size limits $s_{blow}$ and $s_{blow2}$.}
\label{psdsunhot}
\end{figure}

\begin{figure*}
\makebox[\textwidth]{
\includegraphics[scale=0.5]{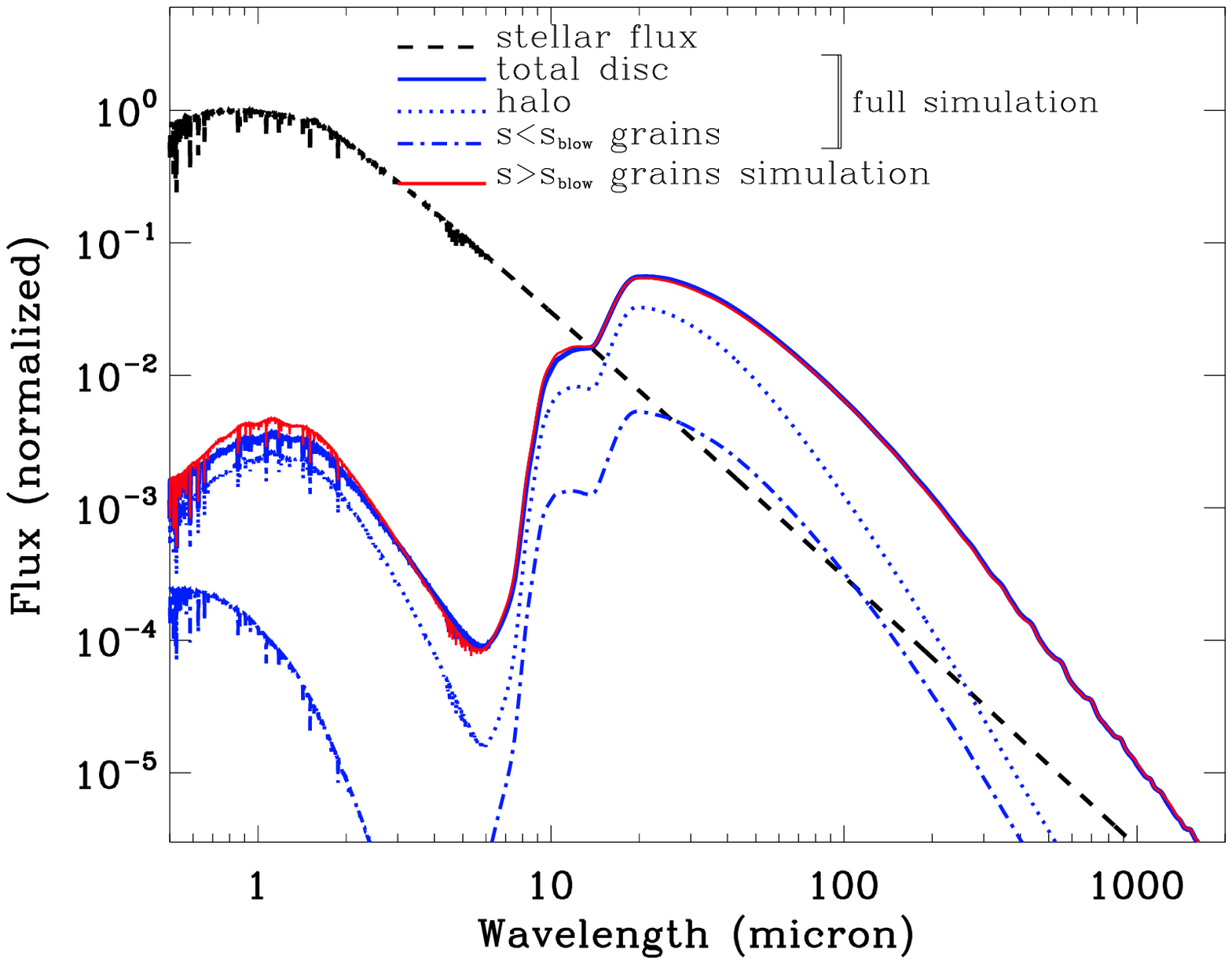}
\includegraphics[scale=0.5]{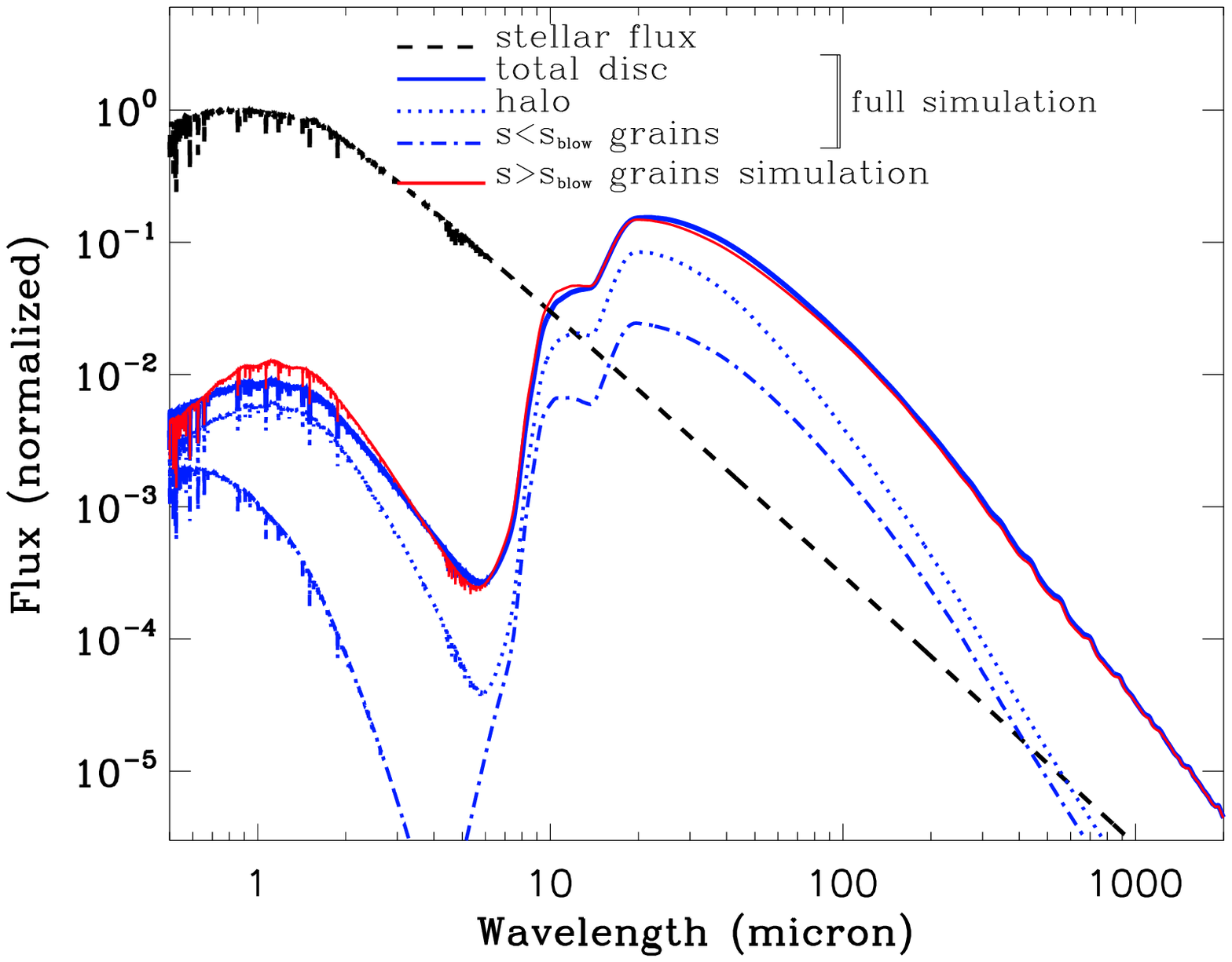}
}
\caption[]{Same as Fig.\ref{sedbphotsil}, but for a solar-type G2V star with $f_d=10^{-3}$ (left panel) and $f_d=5\times10^{-3}$ (right panel).}
\label{sedsunhot}
\end{figure*}


The situation does not radically change for a warm asteroid-like belt. The only difference is that, because of higher impact velocities, the erosion of small bound grains by the tiniest particles is now noticeable on the PSD (Fig.\ref{psdsunhot}). However, the effect on the SED remains relatively limited and only shows up as a scattered-light flux deficit (the missing small bound grains' contributions being here, contrary to the A-star case, not compensated by the contributions of the unbound grains), and a flatter SED in the $\sim0.7-2\mu$m domain.

\section{Discussion} \label{discu}

Our simulations have shown that bright debris discs (with $f_d\gtrsim10^{-3}$) at collisional steady state are able to "naturally" produce a significant amount of small $s\leq s_{\rm{blow}}$ unbound grains, without the need to invoke transient and/or violent processes such as the breakup of large planetesimals or collisional avalanches. In addition, these grains can also significantly erode the population of small bound grains in the $s_{\rm{blow}}\leq s\lesssim$2--3$s_{\rm{blow}}$ domain.

For solar-type stars, $s\leq s_{\rm{blow}}$ grains can account for at least 90\% of the disc's geometrical cross section, but leave a marginal signature on the SED. For bright A-type stars, however, these unbound grains, even if they contribute less to the total cross section, can leave a much more significant imprint on the SED. This is especially true in two wavelengths domains: in scattered light in the visible to near-IR ($0.5\lesssim \lambda \lesssim2\mu$m), and in thermal emission in the mid-IR  ($10\lesssim \lambda \lesssim20-30\mu$m).

We now discuss to what extent these results can be compared to observational constraints. Exploring the specificities of several individual debris discs would exceed the scope of this paper, and we will instead investigate if our simulations can explain some general characteristics or trends that have been observationally derived for bright debris discs. We will distinguish two main categories of observationally-derived features:  those that have been attributed to sub-micron grains without a satisfying physical explanation for their presence or with explanations involving transient events (blue colour, silicate feature) and features that have so far been explained without resorting to blow-out grains but might be, at least in some cases, attributed to them when considering the present results (high mid-IR brightness, double-belts).

\subsection{mid-IR flux}

Our results show that, for an A-star, $s\leq s_{\rm{blow}}$ grains dominate the flux in the 10--20$\mu$m wavelength domain for bright discs with $f_d\geq10^{-3}$. This is a relatively counter-intuitive result, since these wavelengths are much larger than $2\pi s$.
For warm discs ($5\leq r\leq9\,$AU) this is primarily because these submicron grains dominate the total geometrical section $\Sigma$ (Fig.\ref{psdbphotsil}), and that this dominance is strong enough to compensate for their lower emissivity at these wavelengths. For cold discs, submicron grains do not dominate $\Sigma$, but their higher temperature ($\sim115\,$K instead of $\sim80\,$K for larger grains) results in an exponential increase of the Planck function (since we are deep in the Wien part of the function) that compensates for both their lower $\Sigma$ and their smaller absorption efficiency $Q_{\rm{abs}}$.  
This added luminosity due to unbound grains also makes the discs appear brighter in the mid-IR, by at least a factor 2, than what would be expected otherwise. This is especially interesting for cold discs, which might become detectable in the mid-IR despite being located far beyond the stellar distances for which the thermal emission is expected to peak around 10--20$\mu$m. This could, for example, provide a "natural" explanation for the observation of a bright mid-IR clump at 50-70\,AU from $\beta$\,Pictoris by \cite{tele05}, especially since this clump seems to contain a significant population of submicron grains.

\subsection{Silicate band at 10$\mu$m} \label{silband}

\begin{figure*}
\makebox[\textwidth]{
\includegraphics[scale=0.5]{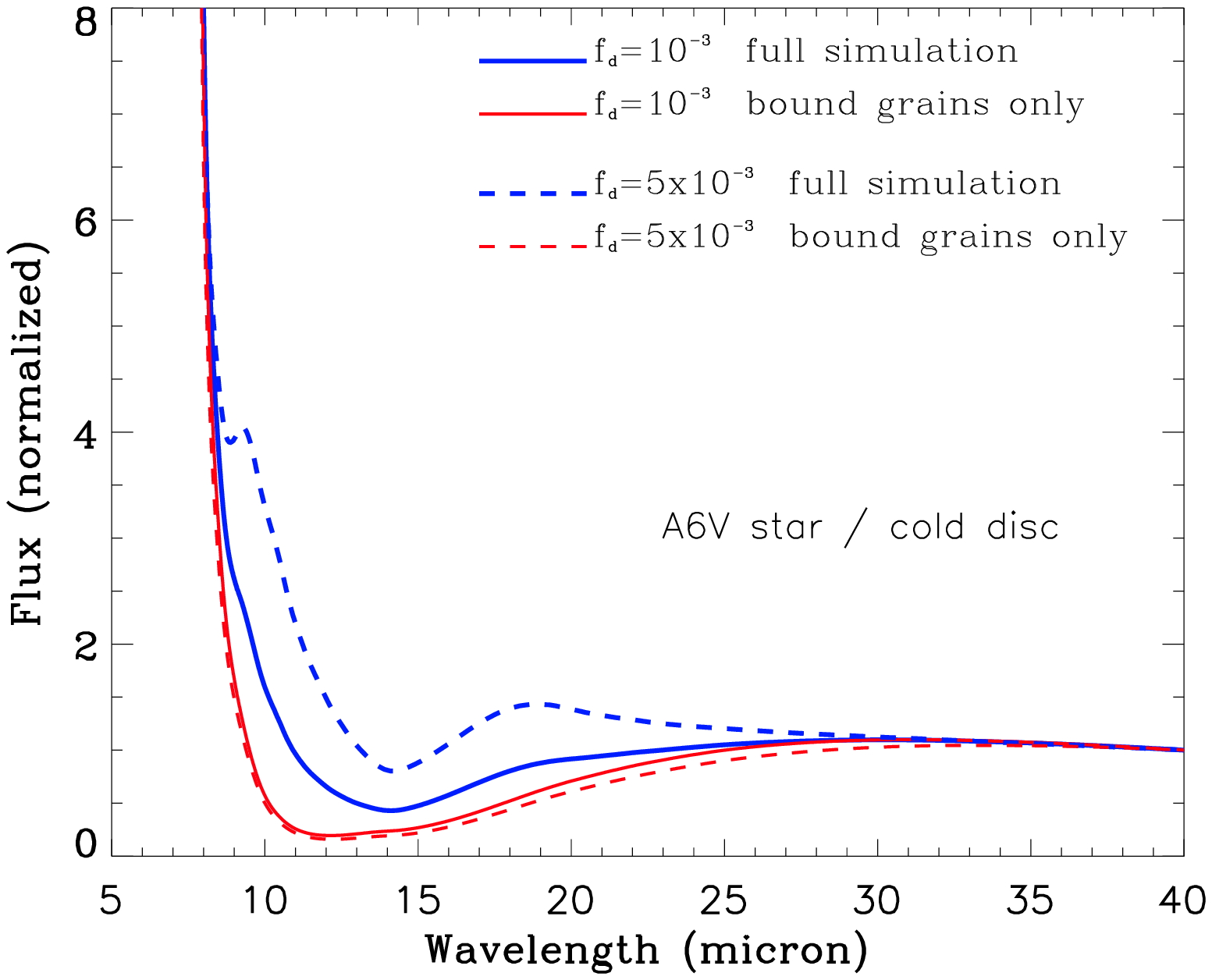}
\includegraphics[scale=0.5]{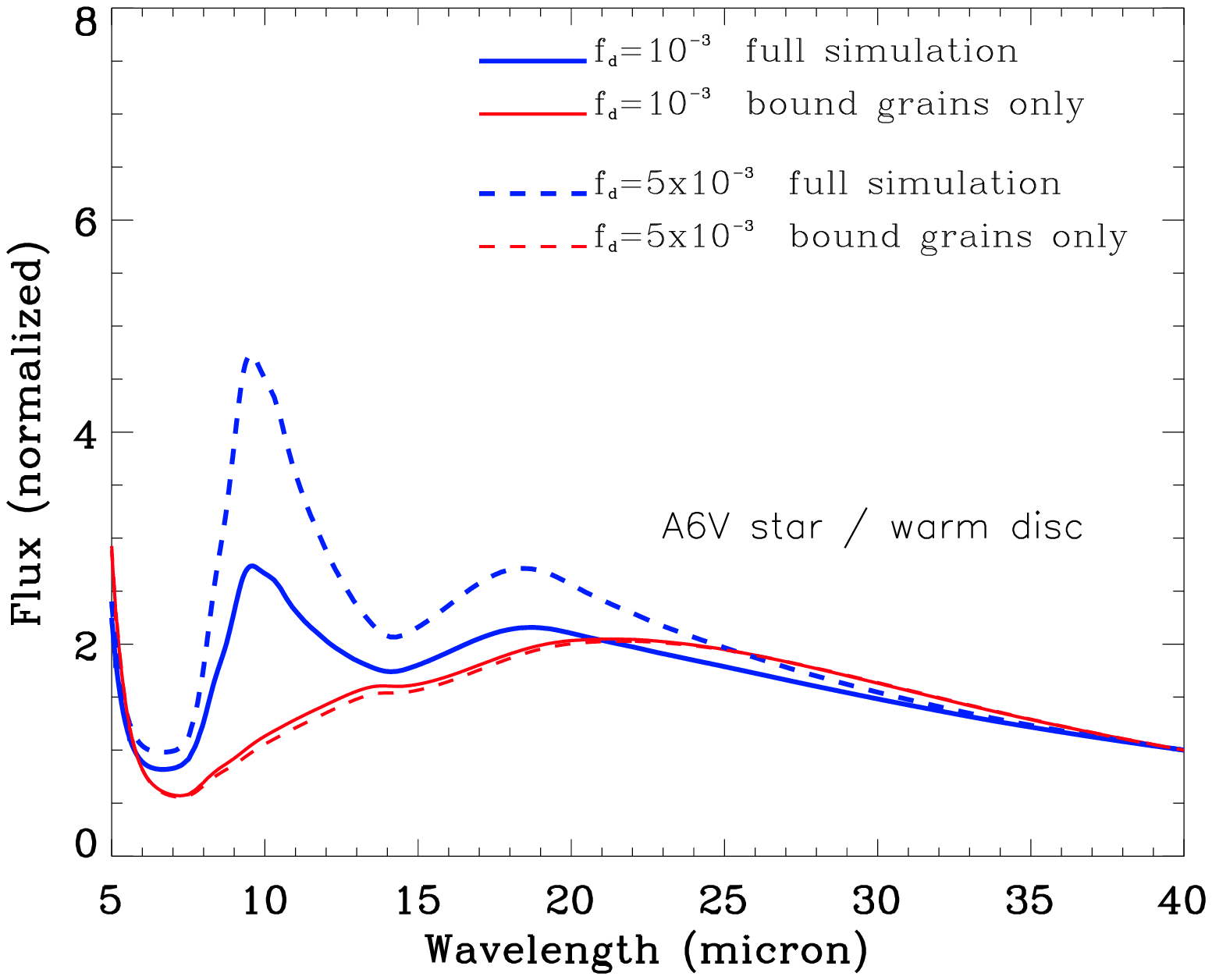}
}
\caption[]{A6V star runs. mid-IR flux divided by the blackbody continuum extrapolated from the flux at $\lambda=40\mu$m assuming $T=115\,$K for the cold disc case, and $T=275\,$K for the warm disc case.}
\label{silline}
\end{figure*}

Another consequence of the important contributions of unbound grains to the mid-IR flux is that the silicate line around 10$\mu$m becomes noticeable in the SED. This is especially true for the warm disc case, where it clearly appears on top of the continuum (Fig.\ref{silline}b). This is not a surprising result in itself since this silicate line has been repeatedly observed in several warm or hot discs and often interpreted as a sign for the presence of submicron silicate grains \citep{matthews2014}, being due to a surge of the absorption coefficient $Q_{\rm{abs}}$ at these wavelengths (attributed to stretching of the Si-O bonds) for grains of sizes $s\lesssim2\mu$m \citep{vosh08}. This has been for instance the case for HD 69830 \citep{beich05,lisse07}, HD32297, HD 113766 \citep{lisse08}, HD 172555 \citep{lisse09,john2012b} or BD+020 307 \citep{wein11}.
However, these observational studies were often not able to put constrains on the quantity of submicron grains associated to the lines, or if they did, it was without exploring possible physical explanations for their presence. We are here for the first time able to show that, for bright warm discs around early-type stars, the steady-state collisional evolution of the system "naturally" produces enough submicron astrosilicate grains for the 10$\mu$m line to appear. Some additional exploratory runs show that the $10\mu$m line remains noticeable down to disc fractional luminosities $f_d\sim5\times10^{-4}$. 

The $10\mu$m feature being due to a variation of $Q_{\rm{abs}}$, it is in principle independent of temperature. As such, it should thus also be detectable in cold discs with significant amounts of submicron grains. However, even when subtracting the continuum, we fail to clearly identify the feature (Fig.\ref{silline}a). This is mainly because scattered light has a dominant contribution to the flux up to the $\sim10\mu$m domain and drowns a large part of the signal due to the silicate emission line. As a matter of fact, the right-hand side of the silicate line is visible on Fig.\ref{silline}a, but quickly merges into the scattered light flux at lower wavelengths. Another factor that makes the $10\mu$m line very difficult to detect in cold disc observations is that it corresponds to a wavelength domain where the thermal emission of these distant discs is very low. In addition, even if the thermal flux is detected around $\lambda\sim10\mu$m, its very sharp variation with $\lambda$ makes the silicate feature very difficult to detect in practice. Last but not least, the contribution from even a small amount of hotter dust located closer to the star would completely drown, at $10\mu$m, the signal coming from these outer regions.
To our knowledge, the only system where the 10$\mu$m silicate band has been identified in a cold ($T\lesssim200\,$K) belt is HD15745, which \citet{mitt15} fitted with grains at $T=156\,$K.

For warm discs around solar-type stars, the silicate line is clearly visible in our synthetic spectra, even before subtracting the continuum (Fig.\ref{sedsunhot}). However, small unbound grains only marginally contribute to it. This is mainly because the blow-out size $s_{\rm{blow}}$ is already well into the $s\lesssim2\mu$m \ domain, so that the abundant bound grains close to $\sim s_{\rm{blow}}$ can produce a strong silicate feature. This result is in agreement with the fact that the 10$\mu$m silicate feature is more readily observed around G-stars than earlier-type stars \citep{matthews2014}.

We note, however, that the observation of a clear silicate feature at $\lambda\sim10\mu$m is not always interpreted as a tell-tale sign for the presence of submicron grains. The strength, peak position and shape of the silicate band is indeed sensitive to the exact physical compositions of the grains, such as the mix between different types of silicates, mainly olivine (Mg$_{2x}$Fe$_{2-2x}$SiO$_{4}$) or pyroxene (Mg$_{x}$Fe$_{1-x}$SiO$_{3}$), the fraction $x$ of Mg and Fe in their composition, or the balance between amorphous and crystaline components \citep{drai03,henning10,olof10}. This probably explains why some studies have been able to fit the observed mid-IR spectra (including the 10$\mu$m feature) of several discs with silicate grains having a particle size distribution (PSD) not extending far below the blow-out size $s_{\rm{blow}}$ \citep[e.g.][]{ball14,mitt15}. However, it is important to stress that these fitting procedures relied on very simplified assumptions for the PSD, notably the absence of any discontinuity at the $s=s_{\rm{blow}}$ limit, which made it very difficult to take into account the potential effect of unbound grains in a reliable way.

\subsection{Colour index}

\begin{figure*}
\makebox[\textwidth]{
\includegraphics[scale=0.5]{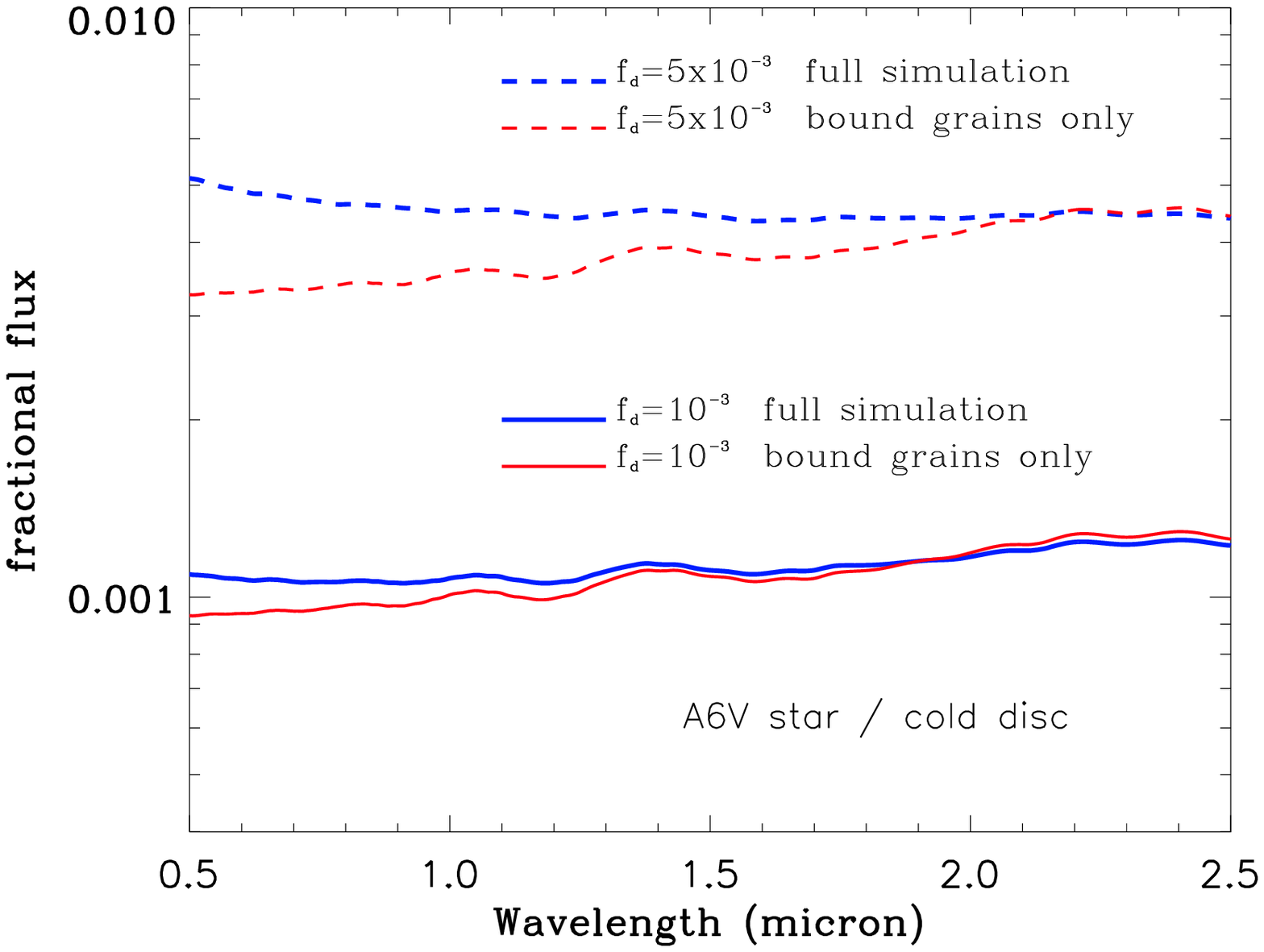}
\includegraphics[scale=0.5]{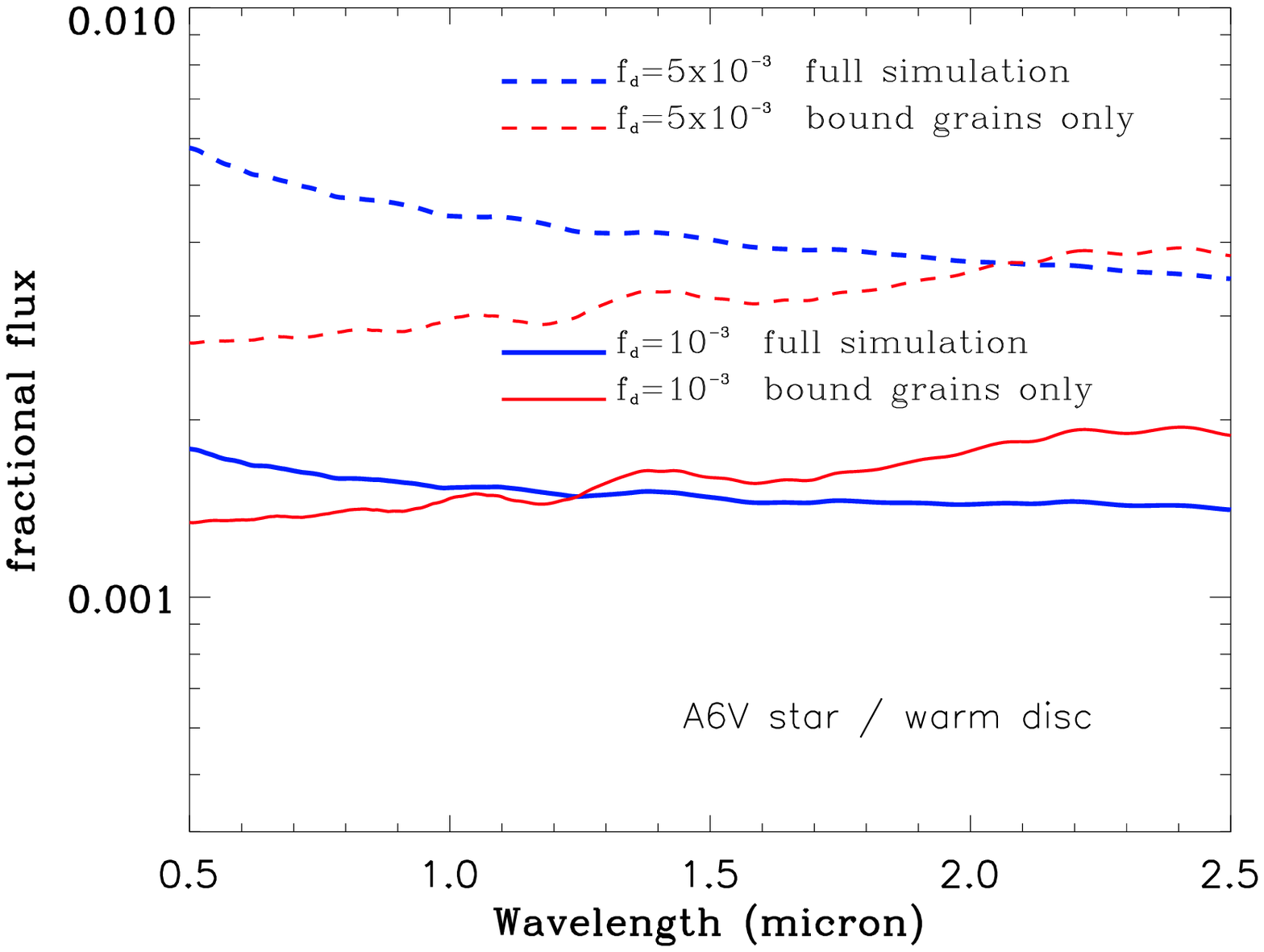}
}
\caption[]{A6V star runs: Ratio of the disc's scattered light flux to the stellar flux in the near-IR.}
\label{colorbp}
\end{figure*}
\begin{figure*}
\makebox[\textwidth]{
\includegraphics[scale=0.5]{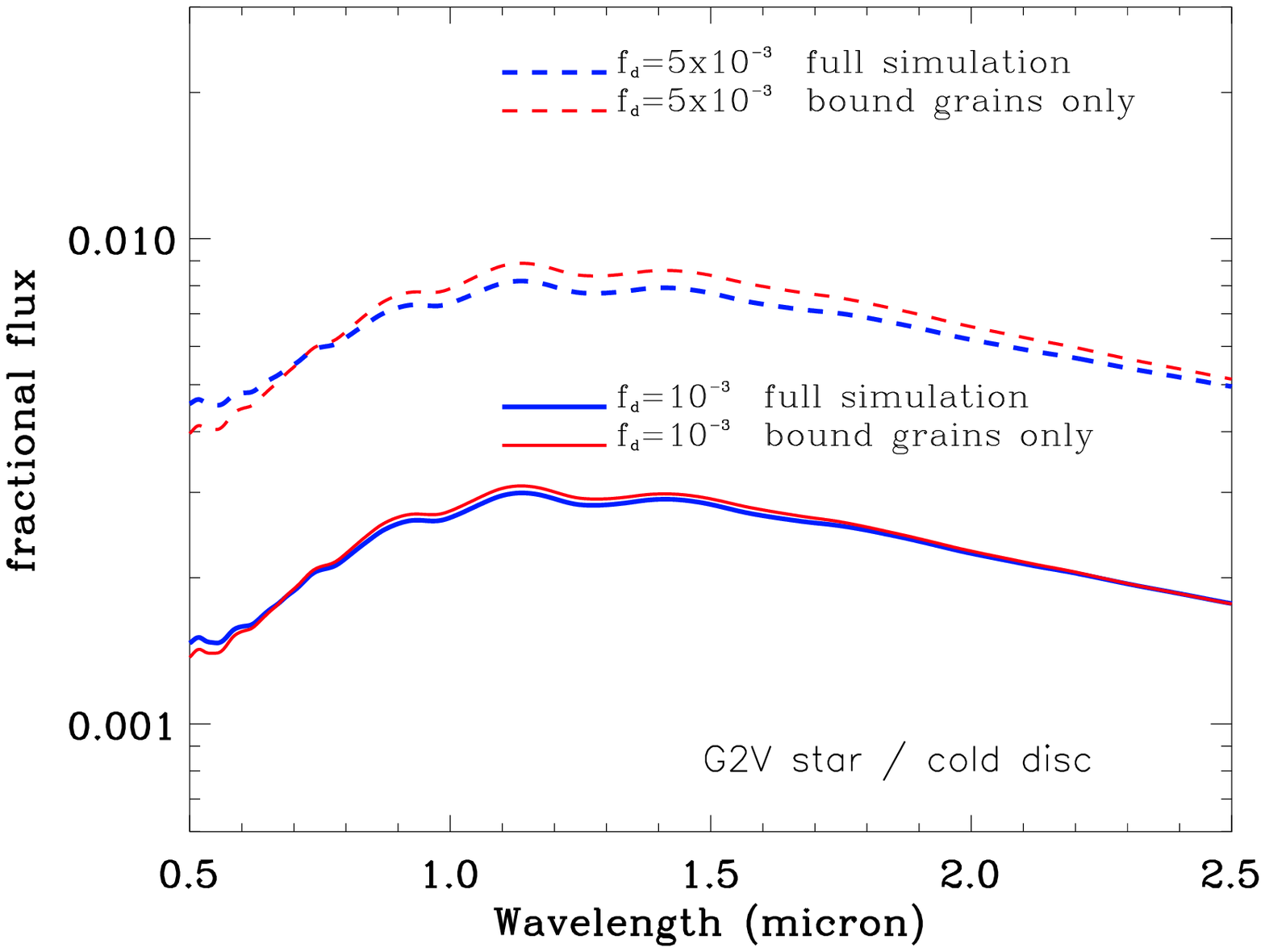}
\includegraphics[scale=0.5]{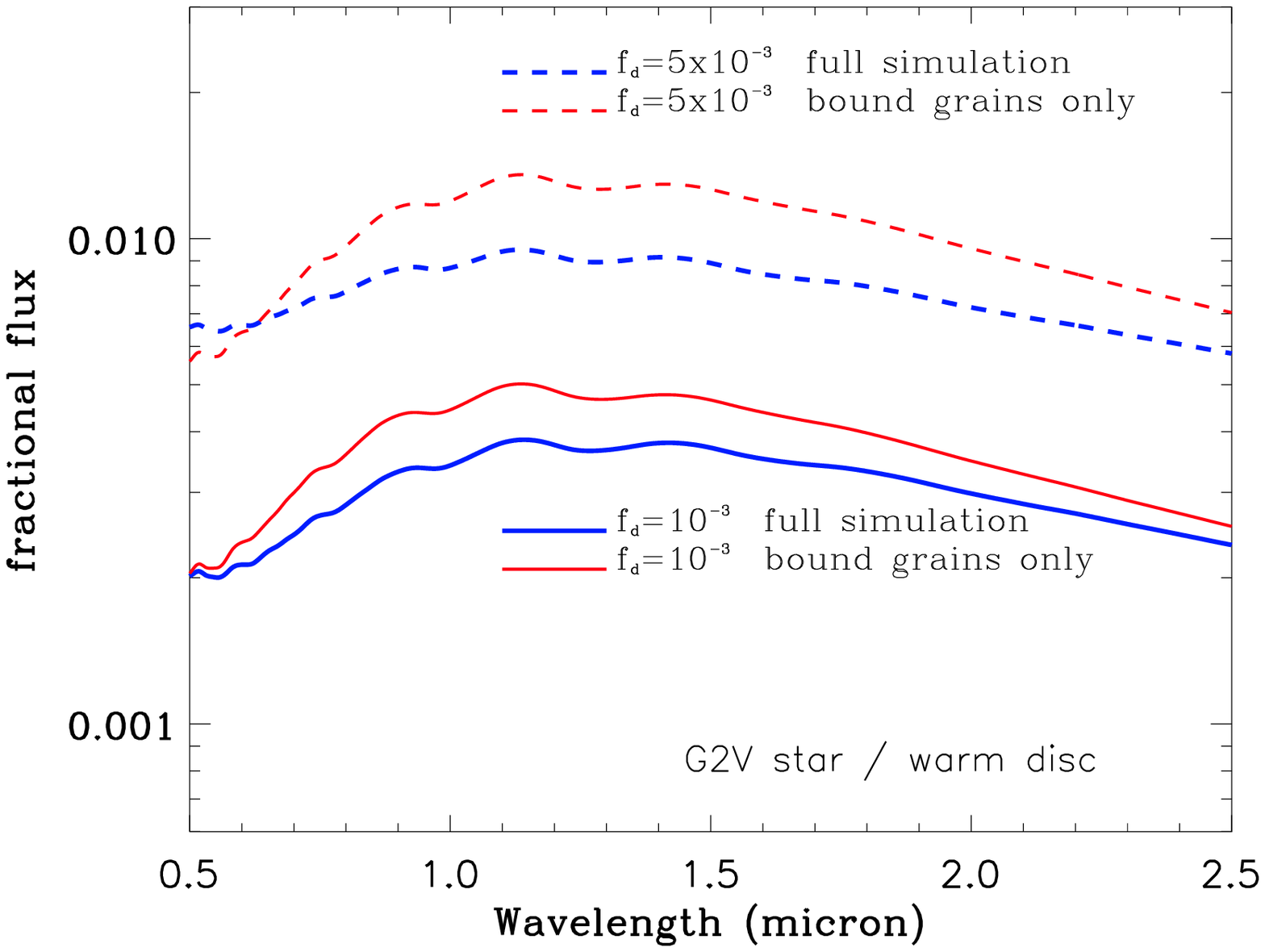}
}
\caption[]{G2V star runs: Ratio of the disc's scattered light flux to the stellar flux in the near-IR.}
\label{colorsun}
\end{figure*}

Another observationally derived tell-tale sign of the presence of submicron grains is a blue color of the spectra, when divided by the stellar contribution, in the visible-to-near-IR range. It is indeed expected that very small sub-micron grains (bound or unbound) Rayleigh scatter and are therefore blue while grains larger than a few micron scatter neutrally, and grains in between would appear red \citep[e.g.][]{meyer07}. A handful of systems with a blue scattered-light color are known to date, such as HD 32297 \citep{kalas05}, HD 61005 \citep{espo16}, the blue needle HD 15115 \citep{debes08}, and the famous AU Mic \citep{lomax18}.
As with the 10$\mu$m silicate band, the amount of submicron grains required to explain the observed blue colours is often not quantified or, if it is, it is by using radiative transfer models assuming unrealistically continuous particle size distributions smoothly crossing the $s_{\rm{blow}}$ boundary. In most cases, the level of this unbound grain contribution is identified as a puzzling feature, interpreted as the consequence of violent and/or transient events (with the notable exception of \cite{fitz07a}, who mentioned the potential role of grain composition and/or porosity).

We show, for the first time, that, for A-type stars, a bright debris disc at collisional steady state can "naturally" produce enough submicron grains to induce a blue colour in scattered light. This is clearly illustrated in Figs.\ref{colorbp} and \ref{colorsun} showing the slope of the fractional $F_{disc}/F_*$ luminosity in the 0.5-2.5$\mu$m range. For all A star cases except that of the cold disc with $f_d=10^{-3}$, the scattered-light spectrum turns from red to blue when taking into account the effect of unbound grains. And even in the remaining cold disc with $f_d=10^{-3}$ the disc is still significantly bluer than in a run stopping at the $s=s_{\rm{blow}}$ boundary. We note that, among the few systems with observed blue colors, two (HD32297 and HD15115) correspond to bright discs ($f_d\geq10^{-3}$) around A  or F stars, which can be considered as archetypes of the bright discs modelled in this paper. We also note that both of these discs have been best-fitted when assuming two-belt models, mainly to account for that blue color and extra emission in the mid-IR, which may just be an artifact of single very bright debris discs (see Sect.~4.4). We also note that, amongst the bright debris discs for which a colour index has been derived, there is a clear "non blue" case: HR4796, which has a red colour in the $\lambda\sim$0.5-1.6$\mu$m domain \citep{debes08a}. More scattered light observations of very bright discs are thus needed to confirm this predicted correlation between fractional luminosity and blue colour. These blue discs may, however, on average be harder to detect (even though they have high fractional luminosities) because of their bluer colour, lowering their fluxes at increasing wavelengths, compared to a standard (redder) bright disc.  

\begin{figure}
\includegraphics[scale=0.5]{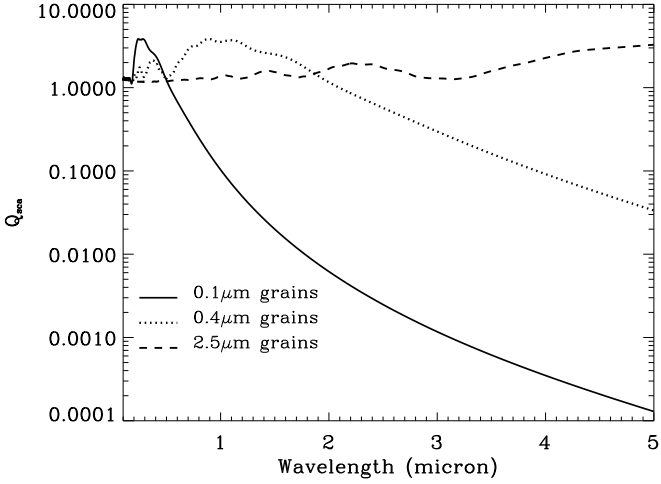}
\caption[]{$Q_{\rm{sca}}(\lambda)$ values for 3 different compact silicate grain sizes}
\label{qsca}
\end{figure}

The effect of $s\leq s_{\rm{blow}}$ grains is less easy to identify for the solar-type cases, because, regardless of the inclusion of $s\leq s_{\rm{blow}}$ grains or not, there is a natural bump in the spectrum around $\lambda\sim 1-1.5\mu$m (see Fig.\ref{qsca}). This bump is due to the $s\sim s_{\rm{blow}}=0.4\mu$m grains that dominate the flux for a G2V star, and whose scattering coefficient reaches a peak value close to $Q_{\rm{sca}}\sim4$ around $\lambda\sim 1\mu$m (Fig.\ref{qsca})  \footnote{This comes from the well known result that grains of a given size are very efficient scatterers at wavelength slightly larger than their size, for which they can scatter a higher amount of electromagnetic energy flux than the one flowing through their geometric cross section \citep{zubk13,kenn15}}. As a consequence, the scattered-light spectrum of a disc around a solar-type star is naturally red in the $\lambda\sim$0.5-1$\mu$m domain and blue in the $\lambda\sim$1-2.5$\mu$m range. The presence of unbound grains is not able to reverse these two slopes, but it can significantly flatten them, especially for very bright warm asteroid-belt-like discs as illustrated in the right panel of Fig.\ref{colorsun}.

 In this exploratory work, we have neglected the potential role of the scattering phase function (SPF) on disc colours, and have considered isotropic scattering for all grains. This issue, as well as other issues that exceed the scope of the present paper (see Sec.\ref{ccl}), will be explored in future studies. However, we expect that the SPF will not radically change our main conclusions on disc colours. Firstly, pronounced anisotropic effects occur only for discs viewed close to edge-on \citep{muld13}, which only represent a fraction of possible viewing angles, and it will occur only for the innermost projected separations. And secondly, the anisotropies of the SPF are expected to be maximum for grains much larger than $\lambda/2\pi$ \citep{muld13,milli17}, whereas the colour effect we focus on here is mainly due to submicron grains that do not meet this criterion.

\subsection{Double belts?}

\begin{figure}
\includegraphics[scale=0.5]{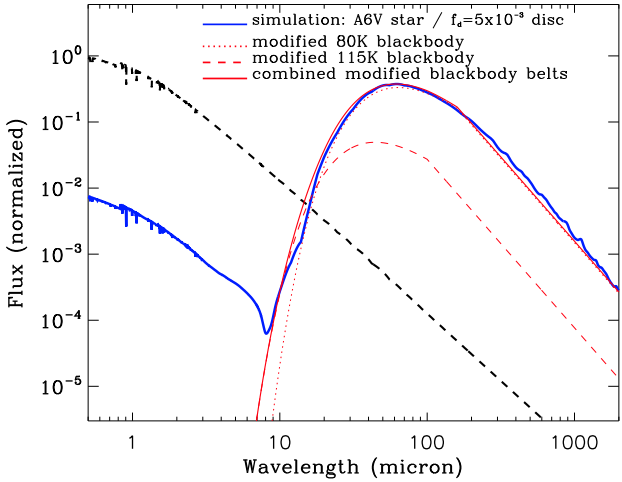}
\caption[]{Comparison between synthetic SED from simulation results (for an A6V star and a cold disc with $f_d=5\times10^{-3}$) and a fit with the combination of two modified blackbodies with T=115K, $\lambda_0=100\mu$m, $\gamma=-0.60$ and T=80K, $\lambda_0=160\mu$m, $\gamma=-0.57$, respectively (see main body of the text for more detai)}
\label{sedmod}
\end{figure}

We find that bright discs around A-stars naturally produce a substancial amount of submicron grains that boosts the signal in the $\sim$10--20$\mu$m wavelength region. Fig.\ref{sedmod} shows that this excess mid-IR flux might mimic the emission from an additional warmer belt instead, and that our synthetic SED for the "very bright" disc case can be reasonably well fitted by the combination of two modified blackbody curves, where the Rayleigh-Jeans part of the spectrum is corrected by a factor $(\lambda/\lambda_0)^{\gamma}$ for $\lambda\geq\lambda_0$, with T=80K and T=115K, respectively. This effect might, however, not be sufficient to explain most of the "double-belts", or more exactly double-temperature configurations that have been inferred for a large number of debris discs \citep{kenn14,geil17}. In most cases, the derived temperature ratio $R_T$ between the two potential components of these systems is indeed of the order of a factor 2--4 \citep{kenn14}, whereas, even for our most "extreme" case, the temperature ratio between our 2 "potential belts" is limited to $\sim1.5$ (corresponding to a separation of a factor $\sim(1.5)^{2}=2.25$ in radial distance). Note that the temperature ratio between that of our smallest submicron grains and that of the largest blackbody-like particles is higher than that, closer to a factor 2, but these larger grains only dominate the SED at wavelengths $\gtrsim100\mu$m and cannot shift the temperature of the "cold" component of our double-belt fit down to a blackbody value. As a matter of fact, \citet{kenn14} also explored the possibility that the range of observed temperatures could arise from different grain sizes in a single belt, but also ruled out this scenario on the same grounds that large blackbody grains do not contribute enough to stretch the temperature range up to observed values. Note, however, that this previous analysis did not take into account any contribution from $s\leq s_{\rm{blow}}$ grains.

Our results could, however, be applied to the few unresolved (or poorly resolved) systems for which double-temperatures with a ratio $R_T\leq2$ have been inferred. According to the "two-grain model" fit by \cite{mitt15} there are 5 systems that potentially match this criterion: HD15745 ($R_T=1.79$), HD23267 ($R_T=1.88$), HD108904 ($R_T=1.56$), HD120326 ($R_T=2.02$) and HD147010 ($R_T=1.88$).

\subsection{The importance of halos}

A peripheral yet important result derived from our simulations is the important contribution that the halo, defined as all the matter that populate the region extending beyond  the outer edge of the main belt, has to the total luminosity of the system. Contrary to the other results highlighted in this work, this contribution is, to a first order, independent of the disc's brightness or density (expressed by $f_d$). This can be understood by the fact that the luminosity of this halo is, for all explored cases, dominated by small \emph{bound} grains, with unbound particles only making a minor contribution. Since the number of these small bound grains is directly proportional to the density in the main disc (because they are collisionally produced and destroyed in this disc), their relative contribution compared to that of the main disc remains constant regardless of the disc's density (for a more thorough discussion on the production and destruction of halo grains, see \cite{stru06} and \cite{theb08}. 

With the exception of a narrow dip at the scattered-light/thermal-emission transition, we find that, for all considered cases, the halo contributes to roughly $\sim$50\% of the flux up to $\lambda$$\sim$50$\mu$m (see Figs.\ref{fracbpcoldsil} \& \ref{fracbphotsil}). Its contribution decreases at longer wavelengths because the small grains it is made of become poor emitters at wavelengths much longer than their size. These results have important consequences for observationally unresolved systems, for which we conclude that all photometric fits have to take into account the contribution of the halo.

\section{Conclusions and Perspectives}\label{ccl}

The observed characteristics of some specific debris discs point towards the presence of submicron grains, which seems counter-intuitive given that such grains should be blown out by radiation pressure on short timescales. As a consequence, this presence of small unbound $s\leq s_{\rm{blow}}$ dust is often interpreted as the result of some transient and/or violent event.

We revisit this issue by exploring to what extent the presence of these small grains could be explained by the "natural" collisional evolution of a bright debris disc with a high collisional activity. To this effect, we use a state of the art numerical code to estimate the fraction of $s\leq s_{\rm{blow}}$ grains in bright debris discs at collisional steady state. We consider two different stellar types (A6V and G2V), two different disc types ("warm" asteroid-like and "cold" Kuiper-belt-like), and two different disc brightnesses ("bright" disc with a fractional luminosity $f_d=10^{-3}$, and "very bright" disc with $f_d=5\times10^{-3}$. 

For A stars, we find that small unbound grains always leave a detectable signature in bright discs with $f_d\gtrsim10^{-3}$. In addition to their own contribution to the disc's luminosity, the grains are also able to efficiently erode the smallest bound grains with $s\geq s_{\rm{blow}}$. The effect of unbound grains is most noticeable in two distinct locations of the disc's SED: in scattered light in the optical to near-IR domain ($0.5\lesssim \lambda \lesssim3\mu$m), and in thermal emission in the $10\lesssim \lambda \lesssim20\mu$m region. In scattered light, small unbound grains are able to switch the disc's color from red to blue. In the mid-IR, the higher temperature of $s\leq s_{\rm{blow}}$ grains compensates for their lower emissivity, an effect that is at its maximum for cold discs where we are in the Wien domain of the Planck function. In the  $10\lesssim \lambda \lesssim20\mu$m range, submicron grains boost the disc's luminosity by at least a factor of 2, which might allow some systems (especially those with cold discs) to become detectable at these wavelengths. In addition, the silicate solid-state feature at $\sim10\mu$m also becomes clearly noticeable, albeit this time only in the case of warm discs. We also find that the additional luminosity contribution due to unbound grains can mimic the SED of two debris belts separated by a factor of $\sim2$ in radial distance, although the effect might not be powerful enough to explain the majority of "double-belts" systems that have been inferred from observations, for which the expected separation between the belts is closer to a factor 5--10.

For G stars, $s\leq s_{\rm{blow}}$ grains can account for more than 90\% of the disc's total geometrical cross section. However, their effect on the disc's luminosity remains limited. This is mainly because the $s_{\rm{blow}}\sim0.4\mu$m limit is itself in the submicron range, so that the  $s\leq s_{\rm{blow}}$ domain is confined to extremely small grains whose higher-temperature can no longer mitigate their low emissivity. The $\sim10\mu$m silicate band is clearly visible, but it is mostly due to the smallest grains on bound orbits just above the $s_{\rm{blow}}$ limit. As for the disc's color in scattered light, the contribution of $s\leq s_{\rm{blow}}$ grains is able to flatten the natural "bell shape" of the spectrum in the 0.5--2.5$\mu$m range that is due to the peak in scattering efficiency around $\lambda\sim1\mu$m of the $s\sim s_{\rm{blow}}$ grains that dominate the flux.

With the data at hands, it is too early to test all of our predictions, such as identifying a correlation between high fractional luminosity discs and blue colours, or the possibility of detecting cold bright discs with silicate features, because of small statistics issues. However, the JWST will be a powerful ally in the close future, both by potentially discovering more blue discs thanks to the NIRCam instrument, and by increasing the number of cold and warm discs with detected silicate features (with MIRI). We note that our results predict that bright discs around A or early-type stars will also be brighter than previously expected in the mid-IR (owing to the additional flux coming from the excess of sub-micron grains "naturally" produced in these discs), therefore improving the odds of detecting debris discs in the mid-IR with the JWST.

There are of course some issues that exceed the scope of the present exploratory study, whose main purpose was to identify a potentially important, and so far neglected consequence of collisional steady-state in debris discs. Such issues, such as a thorough investigation of the role of grain physical composition, of the scattering phase function or the potentially important role of stellar magnetic field and stellar wind on submicron grains, are deferred to future work, which will also explore the specific cases of some individual debris discs. 

\begin{acknowledgements}
The authors warmly thank Julien Milli, Grant Kennedy and Jean-Charles Augereau for enlightening discussions.

\end{acknowledgements}

{}

\clearpage

\end{document}